# Critical Evaluation of Artificial Intelligence as Digital Twin of Pathologist for Prostate Cancer Pathology

Okyaz Eminaga, MD/PhD [1,2], Mahmoud Abbas MD *[3], Christian Kunder, MD/PhD *[4], Yuri Tolkach, MD *[5], Ryan Han, MS [6], James D. Brooks, MD [1], Rosalie Nolley, MS [1], Axel Semjonow, MD [7], Martin Boegemann, MD [7], Robert West, MD/PhD [4], Jin Long, PhD [8], Richard E. Fan, PhD [1] and Olaf Bettendorf, MD[9]

*Contributed equally.

1) Dept. of Urology, Stanford University School of Medicine, Stanford, CA, U.S.A.

2) AI Vobis, Palo Alto, CA, U.S.A.

3) Prostate Center, Dept. of Pathology, University Hospital Muenster, Muenster, Germany

4) Dept. of Pathology, Stanford University School of Medicine, Stanford, U.S.A.

5) Dept. of Pathology, Cologne University Hospital

6) Dept. of Computer Science, Stanford University, Stanford, U.S.A.

7) Prostate Center, Dept. of Urology, University Hospital Muenster, Muenster, Germany

8) Dept. of Pediatrics, Stanford University School of Medicine, Stanford, U.S.A.

9) Institute for Pathology and Cytology, Schuettorf, Germany

**Corresponding author:**
Okyaz Eminaga, M.D./Ph.D.
AI Vobis
Palo Alto, CA 94306
United States

Email: okyaz.eminaga@aivobis.com

# Abstract

Prostate cancer pathology plays a crucial role in clinical management but is time-consuming. Artificial intelligence (AI) shows promise in detecting prostate cancer and grading patterns. We tested an AI-based digital twin of a pathologist, vPatho, on 2,603 histology images of prostate tissue stained with hematoxylin and eosin. We analyzed various factors influencing tumor-grade disagreement between vPatho and six human pathologists. Our results demonstrated that vPatho achieved comparable performance in prostate cancer detection and tumor volume estimation as reported in the literature. Concordance levels between vPatho and human pathologists were examined. Notably, moderate to substantial agreement was observed in identifying complementary histological features such as ductal, cribriform, nerve, blood vessels, and lymph cell infiltrations. However, concordance in tumor grading showed a decline when applied to prostatectomy specimens ($\kappa$ = 0.44) compared to biopsy cores ($\kappa$ = 0.70). Adjusting the decision threshold for the secondary Gleason pattern from 5% to 10% improved the concordance level between pathologists and vPatho for tumor grading on prostatectomy specimens ($\kappa$ from 0.44 to 0.64). Potential causes of grade discordance included the vertical extent of tumors toward the prostate boundary and the proportions of slides with prostate cancer. Gleason pattern 4 was particularly associated with discordance. Notably, grade discordance with vPatho was not specific to any of the six pathologists involved in routine clinical grading. In conclusion, our study highlights the potential utility of AI in developing a digital twin of a pathologist. This approach can help uncover limitations in AI adoption and the current grading system for prostate cancer pathology.

# Introduction

Prostate cancer (PCa) is the most commonly diagnosed cancer in men and one of the most prevalent causes of cancer-related death [1]. PCa is usually diagnosed via prostate needle biopsy and may be followed by radical prostatectomy (total removal of the prostate, seminal vesicles, and surrounding tissues) [2]. The management of patients who undergo prostatectomy requires a reliable histopathological evaluation, including the determination of tumor extent and other cancer-related metrics (particularly grading, staging, and tumor volume) [3]. However, the documentation of the spatial distribution of PCa remains a challenging task since the manual segmentation of cancer and grading on histological slides is time-consuming, particularly for prostatectomy specimens. Additionally, the pathological characterization of prostatectomy specimens or biopsy cores requires extensive histological sampling (e.g., embedding in multiple blocks) for accurate tumor grading, staging, and volume estimation [4,5]. The automated identification and delineation of PCa histology could drastically improve the speed of clinical workflows and provide accurate and detailed documentation for clinical and research usage.

Recent advances in artificial intelligence (AI), especially in digital pathology, have shown great potential for automated cancer detection and tumor grading from histology images [6-15]. Despite having promising results, little is known about how far AI is utilizable as digital twin to accomplish tasks frequently occurring during the clinical routine and research and identify weakness in the current grading system. Therefore, we proposed different test conditions to simulate these tasks to identify the utilization boundary of AI as digital twin for managing PCa pathology.

# Results

**Figure 1** provides a summary for the evaluation results of the digital twin (vPatho) on ten test conditions. Detailed results and the results for tests conditions targeting cancer morphologies (cribriform pattern and ductal morphology) and (mesenchymal tissue structure), tumor precursors (HGPIN) as well as integrating the results from the vPatho assessment into the electronic pathology reports were provided in the **supplementary result section**.

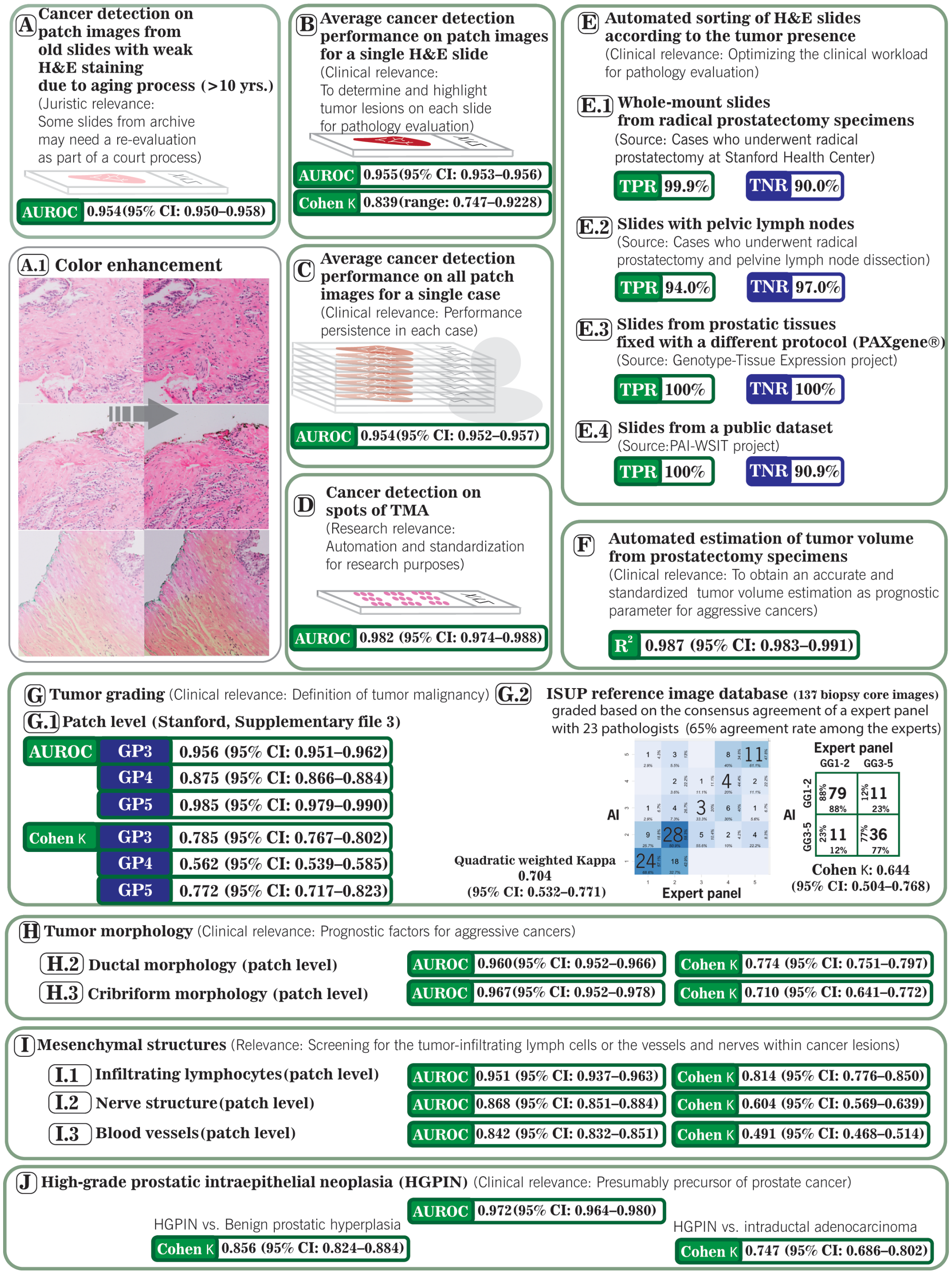

*Figure 1 digital twin's performance under ten test conditions. (A-I) A.1 illustrates the result of the color optimization using a modified version of Macenko. (B) and (C) Whole-mount H&E slide images. (J) This test condition considered all H&E slides from prostatectomy specimens for 136 cases. The relevance for each test is explained. TPR: true positive rate, TNR: true negative rate; AUC: area under the receiver operating characteristic curve; TMA: tissue microarray; H&E: hematoxylin and eosin staining. 95% confidence interval (CI) for uncertainty measurement. GTEx: The Genotype-Tissue Expression (GTEx) project (the images originated from tissues processed using alternative fixatives to formalin; only slides showing preserved glandular morphology are considered for the evaluation and include images with prostatitis). The 11th test condition is provided in the supplementary result section. Results are provided in more details in the supplementary result section.*

**Prostate Cancer**

The concordance level for the delineation of prostate cancer (PCa) areas (defined based on tiled patches) between the pathologist's readings and vPatho was substantial, with a per-slide Cohen Kappa score of 0.8385 (range: 0.7468–0.9284) and per-slide Area Under the Receiver Operating Characteristics (AUROC) of 0.955 (95% Confidence interval - CI-: 0.953–0.956).

The per-slide AUROC was further comparable to the per-case AUROC for PCa detection, emphasizing the performance consistency for the patch-wise tumor detection in each radical prostatectomy specimen when dividing its complete slides into patches.

When we considered images from whole slides (WS) with paled H&E-staining colors archived for more than 20 years, we found that the PCa detection performance on 11,862 patches was comparable to the detection performance on the patches from the recent WS (AUROC: 0.95; 95% CI: 0.950–0.958). Moreover, this comparable performance was achieved by the computationally improved staining condition of these WS images.

Another test condition evaluated vPatho for its accuracy in sorting slides according to the PCa presence status on multiple external datasets. We found that this sorting algorithm delivered excellent sorting accuracy on slide images obtained by different studies, indicating the generalization of vPatho (See Figure 2E).

When we focused on images of the whole mount slides (WM) used for pathology evaluation during clinical routine and sorted them according to the predicted cancer presence status, 99.0% (1,018 of 1,028) of WM histology slides were correctly classified for the presence of PCa; the PPV (positive predictive value) was 99.14%, while the NPV (negative predictive value) was 98.75%. The TPR (true positive rate) was 99.9% and the TNR (true negative rate) was 90.0% (**Figure 1E.1**). Further, a real pathologist would reduce the slide number of normal tissues to examine by 90.0% using vPatho, while missing only a single slide from 1,028 examined slides from 136 cases.

At the case level, we identified 7 erroneous cases per 100 cases. When the false sorting occurred in cases with 8 whole-mount slides on average, at least one slide was falsely sorted. When we looked at the causes for these errors, we found that one of 7 cases had a false negative slide whereas the remaining 6 cases had at least one false positive slide. The error rate for falsely sorted slides in affected cases was 14% (range: 11 – 29%), which is significantly higher than the overall error rate for examined slides (1%).

vPatho could detect lymph nodes with PCa metastases with TPR of 89% and TNR of 97%, although vPatho was neither trained on images from lymph nodes nor on PCa metastases.

The predicted tumor volumes were strongly correlated with the ground truth, with a coefficient of determination ($R^2$) of 0.987 (95% CI: 0.983–0.991) in 46 cases (368 WM histology images) whose histology images had a complete and detailed annotation of the cancerous lesions (**Figure 1F**). **Supplementary result section** provides additional results for cancer presence by vPatho.

## Gleason Patterns and ISUP Grading

In the test condition for the identification of Gleason patterns on images having prostatic tissues in a dimension (~512 $\mu$m) suitable for laser microdissection, we identified that vPatho provided a substantial concordance level with the pathologist's annotation for Gleason pattern 3 (Cohen **κ**: 0.785) and 5 (Cohen **κ**: 0.772). In contrast, a moderate concordance was found for Gleason pattern 4 (Cohen **κ**: 0.562) between vPatho and the pathologist's annotation (See figure 1G.I). Parallelly, we found that Gleason patterns 5 were detected more accurately at 10x objective magnification whereas the detection of Gleason pattern 3 and 4 worked better using patches generated at 20x objective magnification, indicating the performance dependency on the magnification levels for Gleason pattern detection (**See supplementary result section**).

The concordance level for ISUP grade on 137 images from biopsy cores between vPatho and the expert panel with up to 23 pathologists was substantial (quadratic weighted **κ**: 0.70; 95% CI: 0.53 – 0.77). When we binarized the ISUP grades into GG1-2 vs. GG3-5, the consensus rate for GG1-2 (78.2%; 95% CI: 69.0 – 85.8%) between vPatho and the expert panel did not significantly differ from that for GG3-5 (62.1%; 95% CI: 48.3 – 74.5) (P=1.000), indicating no significant preference of our AI algorithms toward any of these subgroups that reflect the malignancy grades of PCa (low-grade vs. high-grade).

On 136 prostatectomy specimens, the concordance level between the pathology reports curated by six different genitourinary pathologists during the clinical routine and vPatho for ISUP grading was moderate (quadratic weighted **κ** = 0.44) before correcting the threshold (5%) for reporting the secondary Gleason pattern. To remind, the primary and secondary Gleason patterns are essential to define the ISUP grade and represent the most frequent Gleason patterns in radical prostatectomy. The definition of the secondary Gleason pattern depends largely on the arbitrary threshold of 5%. If the second most frequent Gleason pattern falls below 5%, the primary Gleason pattern is also considered as secondary Gleason pattern and the second most frequent Gleason pattern is reported as tertiary Gleason pattern.

Given that our previous study on an independent cohort with radical prostatectomy specimens revealed a pathologist-related underestimation of tumor area (underestimation bias) by approximately the half (50%) of the original tumor precentage[4], we corrected this threshold to 10%. Correcting the threshold to 10% significantly improved the concordance level between vPatho and pathology reports from moderate to substantial (quadratic weighted **κ**: 0.64; 95% CI: 0.54 – 0.74). **Figure 2A** shows the confusion matrix for ISUP grading. Here, we found a total of 81 consensus cases (60%) and 55 non-consensus cases (40%).

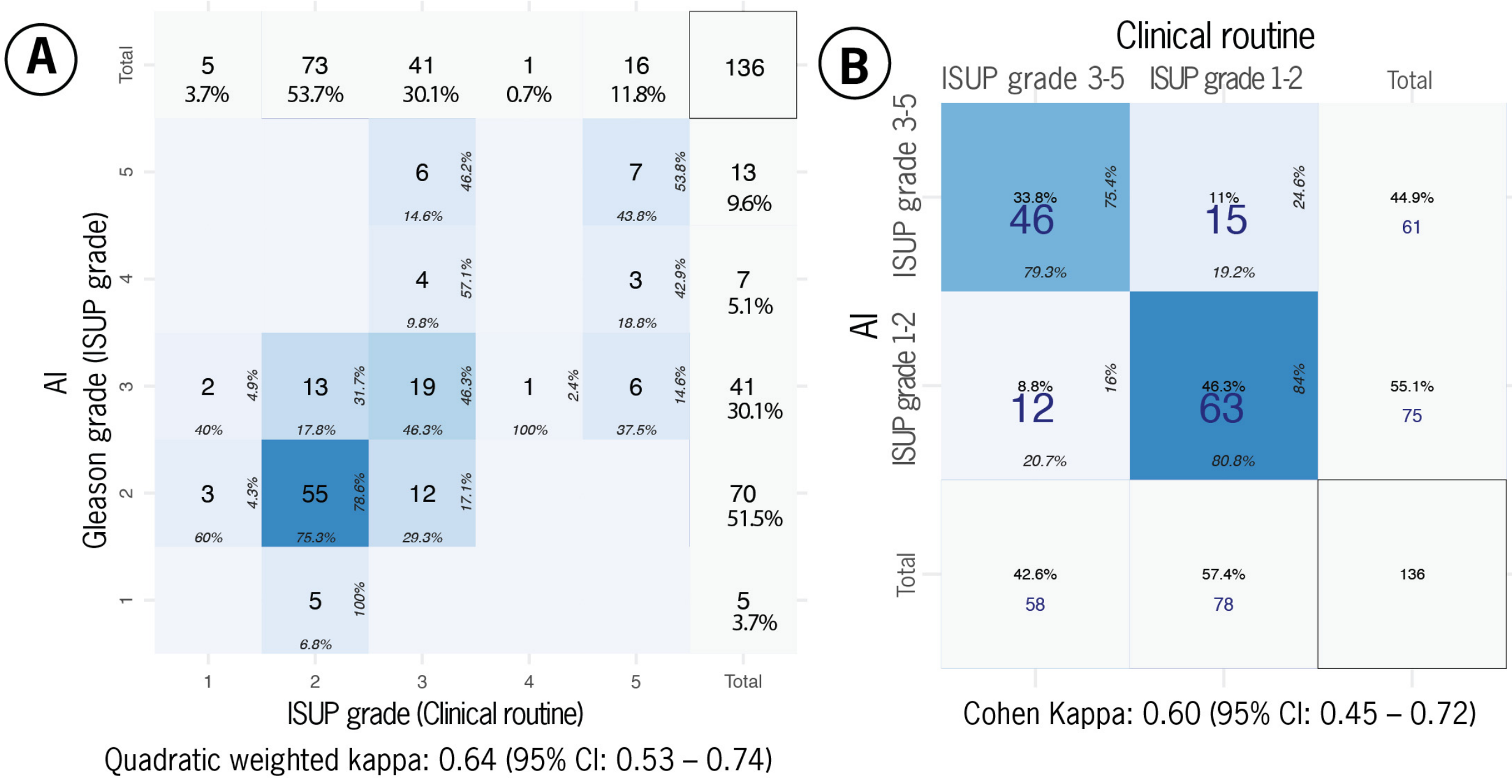

*Figure 2: Confusion matrices for ISUP grades and binary grade groups (ISUP grade groups 1-2 vs. grade groups 3-5) on 136 radical prostatectomy specimens. Kappa values are provided for each endpoint.*

When we divided ISUP grades into GG1-2 and GG3-5 (**Figure 2B**), the consensus rate for GG1-2 (70%; 95% CI: 59 – 79%) between vPatho and pathology reports was comparable to that for GG3-5 (63%; 95% CI: 51 – 74%), indicating no biased preference of vPatho toward to any of these subgroups (P=0.700). We found that 80% of cases (n=109) received the same binarized ISUP grades from vPatho and pathologists, while the remaining 20% cases did not (n=27).

## Factors associated with ISUP Grading discordance

We investigated histopathological factors for their association with the discordance in malignancy grading (i.e., mismatch in ISUP grading between vPatho and pathology report). This evaluation included the pathologists who performed the ISUP grading on prostatectomy specimens, and factors describing the number of slides per case, tumor extension and grading. The factors for tumor extension were the proportion of slides with PCa and tumor volume in percentage (TuVol%). Furthermore, the tumor grades defined by vPatho, and pathologists were considered. Since the complementary version of the Gleason grading system was introduced in 2016 and its adaptation in clinical routine needed sometimes, we included the year of tumor grading as well. For each case, the pathological tumor staging, and the surgical margin status were also collected and incorporated in our analyses. Finally, we conducted mediation analyses to investigate the interaction between the significant indicators for grade disagreement. Detailed statistical results can be obtained in the supplementary section (**See supplementary result section**).

Our analyses identified that ISUP grading made either by pathologists or our AI algorithms was associated with the grade disagreement, when the tumor stage (pT), the locoregional lymph node metastases status (pN) and the surgical margin (R) status were not considered for adjustment of the multivariable models. However, when the multivariable mixed-effects regression model included pT, pN and R status, ISUP grading was no longer a significant indicator for grade disagreement. Interestingly, the capsule proximity as well as the proportion of positive slides were indicative for the grade disagreement, suggesting that other cofounding factors defining the tumor extension direction (i.e., horizontal tumor extension toward the prostate boundary, the number of positive slides for vertical tumor extension) were contributing to the grade concordance status between AI algorithms and pathology reports (**Figure 3 and supplementary result section**). We could not identify individual pathologists as significant indicator for the grade discordance between AI algorithms and the pathology report curated during the clinical routine. The year of grading was weakly associated with the pathologists, but not indicative for the grade disagreement, emphasizing no association between the grade discordance and the adaptation period for the recent ISUP grading system in clinical routine.

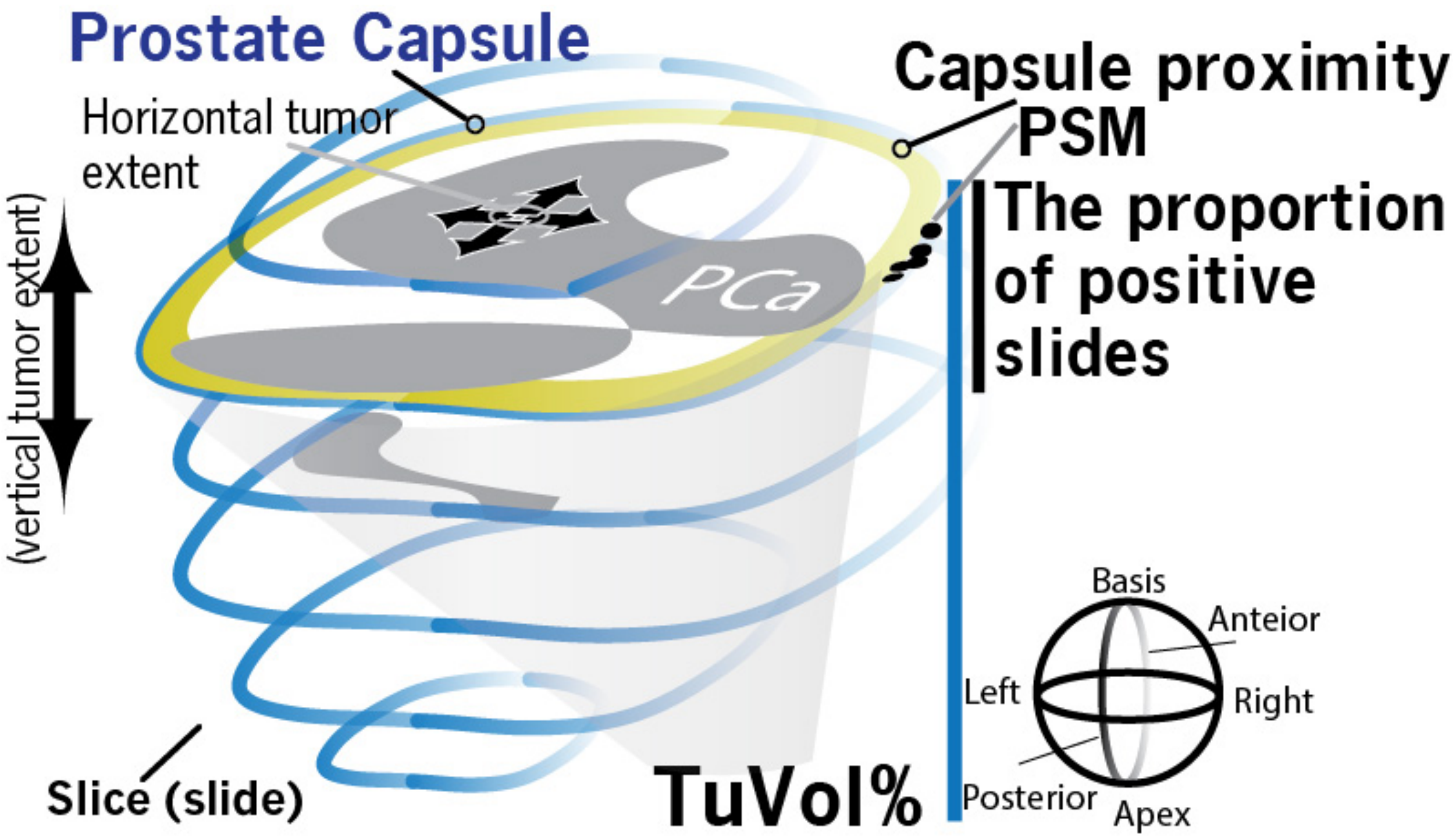

*Figure 3 illustrates a schematic 3D model for prostate with the definition of positive surgical margin (PSM) and the capsule proximity of prostate cancer, the proportion of positive slides, tumor volume in percentage (TuVol%) as well as the definition of horizontal and vertical tumor extents. Each slice here represents a single whole-mount slide. Our analyses with multiple mixed-effects regression models and mediation analyses revealed that the proportion of positive slides for prostate cancer (PCa) and the capsule proximity status of prostate cancer were indicative for the tumor grade concordance level between AI algorithms and pathology reports. While the capsule proximity status was indicative for grade discordance, an increase in the proportion of positive slide was indicative for grade agreement. Because a single radical prostatectomy specimen cannot be embedded into a single block and investigated at once, this specimen is dissected into multiple slices (whole mount). PSM: Positive surgical margins.*

Although TuVol% was associated with the proportion of positive slides and significantly two-fold higher in cases with positive capsule proximity compared to cases with negative capsule proximity, TuVol% was not indicative for the grade concordance levels between AI algorithms and pathology reports, implying that the tumor expansion range (TuVol%) was not enough to trigger the grade disagreement.

Finally, we found that the absolute overall deviation between Gleason patterns 3 (GP3) and 4 (GP4) from the zero point (a point where the percentages of GP3 and GP4 are equal) was similar between cases with grade discordance (34.1%) and those with the grade concordance (35.6%), emphasizing that the absolute deviation from the zero point is not relevant for grade discordance (P= 0.1521). In contrast, the median percentage difference between these Gleason patterns was -16.2% (Interquartile range, IQR: -34.2 – -4.9%) (negative sign: more GP4) for cases with grade discordance and 25.8% (IQR -23.6 – 43.7%) (positive sign: more GP3) for cases with grade agreement, revealing that vPatho estimated the percentage of Gleason pattern 4 significantly higher in the discordance condition compared to those with grade concordance overall (P= 0.0005) or in ISUP grade 2 (3+4) (P=0.01668) as shown in **Figure 4**. Importantly to mention, that the median difference between GP3 and GP4 percentages deviated marginally (by 6.2%) from 10% threshold for cases with grade disagreement, suggesting the existence of a gray zone (uncertainty range) for the definition of grade concordance between vPatho and pathology reports (or a pool of six clinical pathologists) in a subset of these cases. **Figure 5** provides two example cases to stress the impact of the proportion determination in causing the grade discordance between vPatho and the human observer.

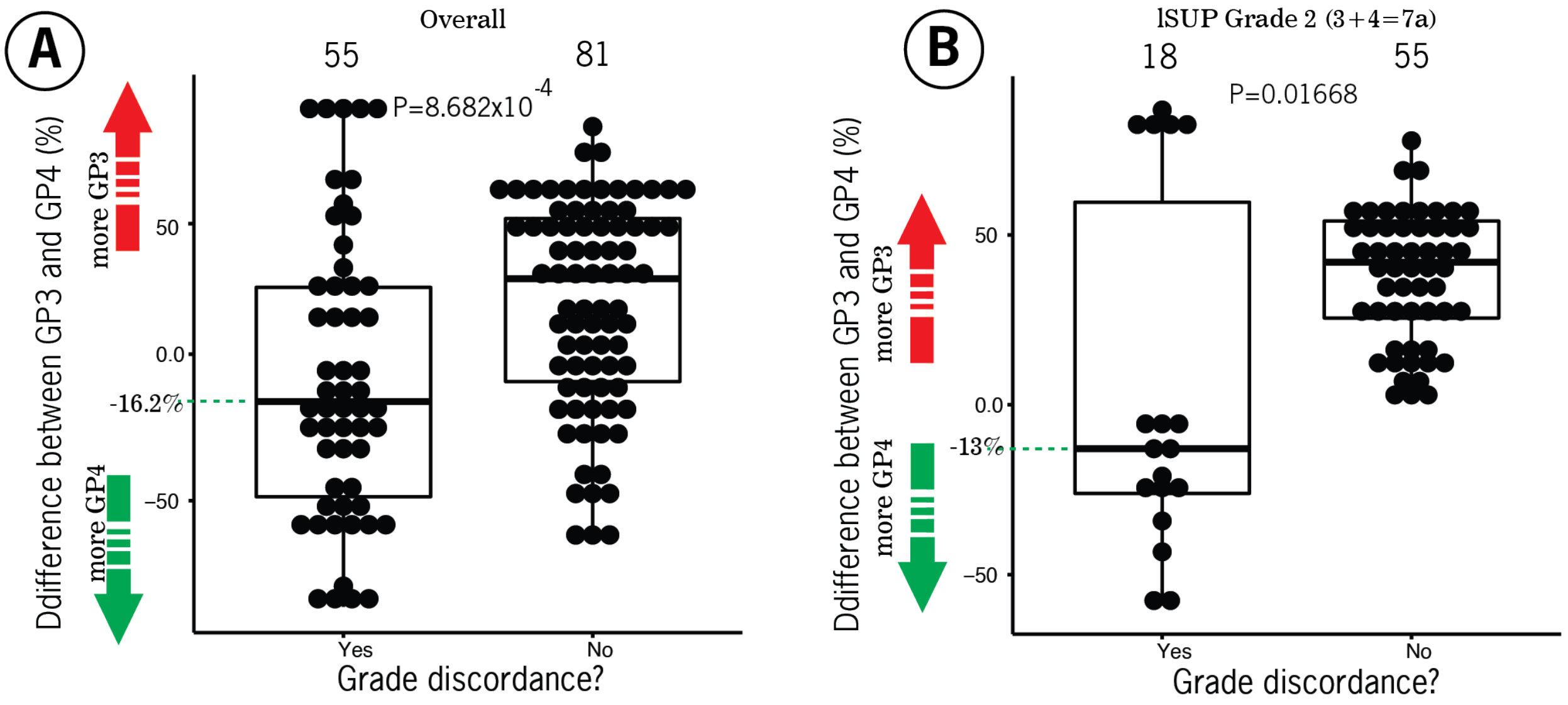

*Figure 4: The box plots describe the proportion differences between Gleason patterns 3 and 4 stratified by the grade discordance status in overall cases (A) and in cases with ISUP grade 2 reported in the pathology report (B).*

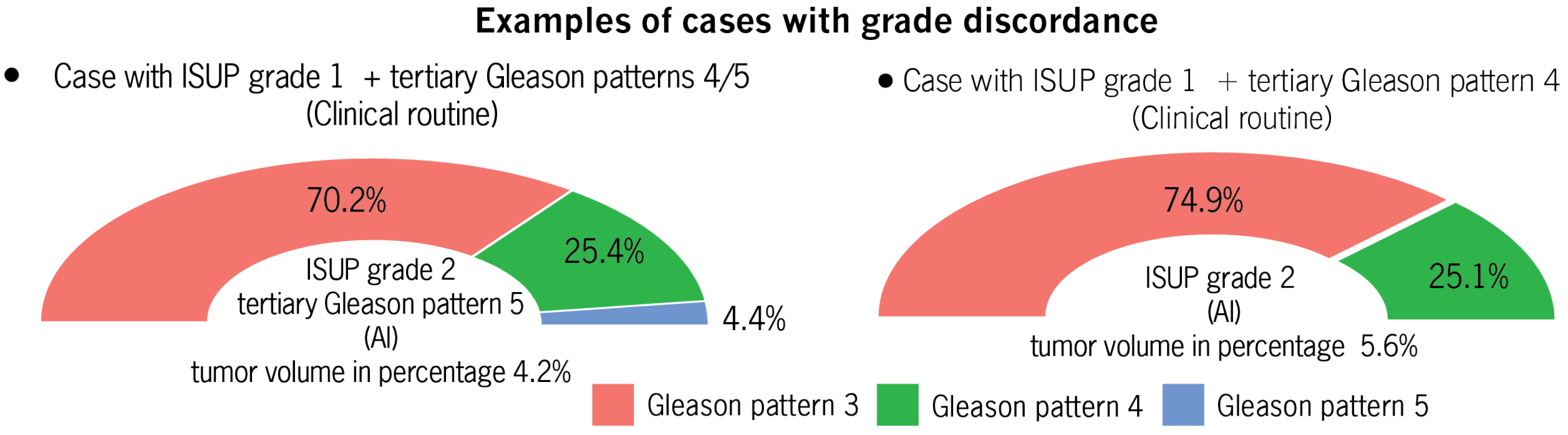

*Figure 5: Two example cases with grade discordance between vPatho and the pathology reports. The proportions are estimated by vPatho (AI) and compared the resulting ISUP grade and tertiary Gleason pattern with pathology reports (Clinical routine). The tumor volume in percentage was calculated by vPatho. Despite both vPatho and pathology report determined the right Gleason patterns in these cases, the difference in ISUP grade occurred. Although Gleason pattern 4 detected by vPatho was affected by a higher false negative rate compared to other Gleason patterns 3 and 5, we found that the grade discordance between vPatho and pathology reports was associated with a higher proportion estimation by vPatho for Gleason pattern 4 compared to Gleason pattern 3 (See **Figure 2**). This finding reveals that the proportion for Gleason pattern marked as tertiary Gleason patterns by pathologists is twice-folder higher than 10% threshold (a corrected threshold for determining the secondary Gleason pattern), indirectly highlighting the inaccurate size estimation (50% of the original size) made by the human observers (pathologists). This result is in accordance with our previous study where we showed that human observers (pathologists) significantly underestimate the size proportion (by 50% of the original tumor) compared to the computer-assisted size estimation on an independent dataset with 255 prostatectomy specimens[4].*

## Discussion

The current study evaluated the digital twin for accomplishing tasks recurrently occurring in the clinical routine to evaluate histology slides with prostatic tissues for PCa. Our findings support the conclusion that building an artificial intelligence (AI) solution that acts as digital twin is promising. Moreover, previous studies showed a potential improvement in concordance level between pathologists using AI [15,16]. However, major challenges remain to realize such AI solutions for clinical stage and are discussed here.

Firstly, the current study reveals a need to develop guidelines for PCa pathology justified to the limitations and the utilization boundaries of AI algorithms. This guideline shall, in addition to image quality assurance [17], incorporate test conditions to ensure a persistent efficacy of AI algorithms on different tissue amounts ranging from micro-dissected tissues to radical prostatectomy specimens, strategies to encounter the AI limitations, as well as the feasibility to integrate the evaluation results from AI algorithms into an electronic pathology report.

Since the definition of expert panel may occur in a closed circle and subject to several biases (e.g., social network bias, implicit bias, situative and geographic bias), we also compared our conclusions with the conclusions of other geographically distinct studies listed in **Table 1**. The concordance levels for vPatho are in alignment with concordance levels reported by various studies for biopsies and radical prostatectomy specimens (**Table 1**), indicating the generalization of our results.

*Table 1 lists the studies examined the recent version of ISUP grade groups and other findings for their concordance level and compared with our concordance levels. The geographic location definition is according to the United Nation geographic Scheme. The list demonstrates that most geographic regions are still underrepresented for the interobserver reproducibility of the recent version of ISUP/ 2016 WHO grading system (e.g., Africa and central Asia), highlighting a significant regional disparity in evaluating the grade discordance. * The concordance level among genitourinary pathologists is higher than the general pathologists. HGPIN stands for high-grade prostatic intraepithelial neoplasia. The conclusions of the current study (7/9) are mostly in agreement with the conclusions of these studies investigated the concordance conditions between different pathologists.*

| Study description | | | | | Concordance level | |
|---|---|---|---|---|---|---|
| **Publication** | **Geographic locations of pathologists** | **Metrics** | **Sample description (Geographic origin)** | **Finding** | **Study's conclusion** | **Our conclusion** |
| Al Nemer et al (2017)[18] | Western Asia | Fleiss kappa | 126 slides with biopsy cores (Western Asia) | ISUP grade group | Substantial | Substantial |
| Dere et al (2020)[19] | Western Asia | Fleiss kappa | 50 biopsy slides from 41 cases (Western Asia) | ISUP grade group | Moderate | Substantial |
| Egevad et al (2018)[20] | Western Europe, North Europe, North America, South America, Eastern Asia, Australia, and New Zealand | Average weighted kappa | 90 core needle biopsies (STHLM3[21], North Europe) | ISUP grade group | Substantial | Substantial |
| Giunchi et al (2017)[22] | Southern Europe | Cohen kappa | 121 regions of interest from 61 slides covering biopsy, radical prostatectomy, and TUR (Southern Europe) | Prostate Cancer | Substantial | Substantial |
| | | | | HGPIN | Substantial* | Substantial |
| van der Slot et al (2020)[23] | Western Europe | Krippendorff's $\alpha$ | 80 radical prostatectomy specimens (Western Europe) | Cribriform pattern | Moderate | Substantial |
| | | | | ISUP grade group | Substantial | Substantial |
| | | | | Gleason pattern 4 | Moderate | Moderate |
| | | | | Gleason pattern 5 | Moderate | Substantial |

Using vPatho has furthermore discovered addressable challenges in the current grading system contributed to the grade discordance in our radical prostatectomy samples, which are:

1. The arbitrary decision threshold for secondary Gleason pattern. The concordance level for radical prostatectomy between vPatho and pathologists (pathology report) was inferior to that for biopsy cores when the decision threshold for secondary Gleason pattern is not adjusted for the underestimation bias. When the threshold is corrected from 5% to 10% based on observation of a previous study[4], the concordance level for ISUP grade groups has significantly improved for radical prostatectomy and consequently eliminated the differences in the concordance levels of ISUP grade groups between biopsy cores and radical prostatectomy specimens. These conclusive findings emphasize the importance of adjusting the decision thresholds for a better consensus level between artificial intelligence and a pool of pathologists or even between pathologists.

2. Distraction of pathologists possibly by attention or cognitive bias directly not related to ISUP grading (e.g., topographical tumor spread) due to the limited human capacity in perception[24]. In contrast, the distraction of AI is related to image quality and content (e.g., brightness, the epithelial of seminal vesicle). Considering distracting factors during model training or image pre-processing can mitigate the distraction effects with accuracy improvement.

3. The proportion of slides with PCa in a single case can impact the grade discordance, possibly due to cognitive bias by human observers[25]. vPatho mitigates the cognitive bias by considering all slides tiled into small patches labelled with PCa for ISUP grading. Moreover, the proportion estimation by a computer-assisted planimetry is more accurate and objective than the human observer's assessment[4].

4. Gray zone for ISUP grade groups due to the proportion closeness can lead to the discordance. Since the computer-assisted planimetry of vPatho has more accurate volume estimation than the human observer, vPatho manages the borderline cases intuitively better than the human observer.

5. The existence of multiple ISUP grade groups is associated with increasing discordance likelihood due to increasing grade complexity. In a simulation with an equal probability for each group, reducing the group number from 5 to 2 improves the concordance likelihood between two pathologists from 4% to 25% just by chance. Moreover, the current study found that the major driver for discordance is Gleason pattern 4 in alignment with two previous studies [4,13]. A possible strategy to increase the concordance level would be weighing the proportion of Gleason pattern 3 in the group definition (e.g., dominant vs non-dominant/rare or absent Gleason pattern 3) instead of Gleason pattern 4 that cover more heterogenous and diverse patterns with a complex and challenging boundary definition.

By analyzing the limitations of the grading system, we identified that these limitations are rooted in the practical recommendations of the current ISUP grading system that heavily depends on the subjective estimation of Gleason patterns for ISUP grading[26]. Moreover, the current grading system is adapted to the pathologists' limitations by providing practical measurements for clinical implementation that are not time- or effort-intensive[27], leading however to an increase in the discordance likelihood[28]. Measurements adjusted to the pathologist's limitations are not always transferrable to AI given their distinctive nature. Therefore, developing a widely acceptable AI solution requires a specific version of the grading system or prostate cancer reporting for AI.

Our study reveals the need and feasibility of developing standardized test conditions reflecting clinical routine practice. An example for that need is that false negative/positive slides per case occurred sporadically and were not concentrated to certain cases, highlighting a significant concern of considering a single slide per case to claim a clinical-grade evaluation. Falsely sorting a single slide leads to an error rate between 11% to 29% in a single case depending on the slide number per case (See Table 25 in addendum table). So, it is clinically more important to report how many cases have falsely sorted slides per 100 cases examined. Furthermore, we emphasize that human examiners require additional efforts to screen for false negative slides (the human examiner needs to screen the whole slide for prostate cancer to confirm or exclude the existence of a false negative slide) than for false positive slides (we require only to examine the regions demarcated by vPatho to determine false positive slides). We emphasize that having a negligible false negative rate is useful to direct the focus of the pathologists on correcting the false positive slides with cancer areas demarcated by vPatho.

It is worth mentioning that the detection performance on patches (the deepest level of evaluation) from tissue microarray (ROCAUC: 0.982; 95% Confidence interval -CI-: 0.974 – 0.988) was significantly higher than on patches from prostatectomy specimens (AUROC: 0.955, 95% CI: 0.953 – 0.956) given that the histologic heterogeneity likelihood is higher in prostatectomy specimens and therefore more challenging compared to smaller tissues (i.e., laser microdissection, TMA spots); the curation of tissue microarrays (TMA) may rely on the random area selection, but it still follows a targeted tissue sampling that reduces the histologic heterogeneity spectrum. Our results also indicate that the detection performance for prostate cancer on TMA is not representative for the detection performance on slides of radical prostatectomy specimens at the patch level. Overall, the tissue dimension and tissue heterogeneity of prostate samples does impact the detection performance. A variety of tissue dimensions should be therefore considered during the performance evaluation for prostate cancer detection to claim generalizable clinical-grade performance. Additionally, we highlight the importance of reporting the image preprocessing steps as the definition of image preprocessing is strongly associated with the detection performance under the test condition and therefore important as deep learning models.

Finally, we further emphasize the importance of designing the test conditions of datasets as completely disjoint from the development set[29]. The study included the slides examined during the clinical routines to generate the pathology reports. Our test design is inspired by previous studies investigated the concordance levels between pathologists for different pathological findings.

This study has limitations. First, this study did not evaluate the correlation of the Gleason grading with survival outcomes given that the follow-up period for an appropriate survival analysis is lesser than 10 years for the WHO 2016 grading system[30]. Secondly, the AI-based digital twin undergoes constant enhancements through iterative improvements in both content and technical aspects. Third, the current study covered only the major spectrums of prostate cancer pathology and the core goal of this study is to evaluate AI under different test conditions reflecting clinical routine and to demonstrate the useability of AI as digital twin to determine issues associated with the current grading system. The future work will focus on automating the tumor staging and pathology description.

## Conclusions

Our digital twin concept facilitates trouble-shooting challenges in digital pathology and clinical practice for prostate cancer pathology.

# Methods

## Image Database

We utilized publicly available histology images of diagnostic slides provided by The Cancer Genome Atlas (TCGA) [31] and images of tissue microarrays from a previous study [6] for model development. For model evaluation under different test conditions (external datasets), we collected a total of 2,540 images of H&E-stained diagnostic histology slides, micrograph, or tissue microarray spots of paraffin-embedded prostate tissues from 709 cases. These slides were obtained by different institutions and scanned using different types of scanners. **Table 2** provides a summary of the image datasets for each test condition.

*Table 2: The description of the cohort utilized to run the test conditions (external validation set). All histology slides were stained with hematoxylin and eosin. HGPIN: High-grade prostatic intraepithelial neoplasia. +the whole slide images were tiled into small image patches ++ Whole slides with a portion of the prostatic slice (2.3 times smaller than the prostatic slice). TMA: tissue microarray. +++Given that a single whole-mount (WM) slice roughly corresponds to 20-30 biopsy cores and the time- and labor-intensive effort for a high precision annotation of the WM images for prostate cancer, we randomly selected 46 cases with a total of 368 WM images (~ 7,360- 11,040 biopsy core images or ~894,240 non-overlapping tiles (Dimension: 512x512 pixels) at 10x and ~1 µm per pixel) for the patch level performance evaluation. * The images were provided by different sites. The negative control groups for the detection of ductal morphology, cribriform pattern, vessels, nerve structure, and HGPIN are already described in the sections below.*

| Test condition | Data description or project source (number of cases), image size | Number of images/patches | Scanner vendor | Objective magnification level |
|---|---|---|---|---|
| **Prostate cancer detection on the H&E slides** | | | | |
| ***Slide condition or sample dimension*** | | | | |
| *Old slides (~20 years)* | | | | |
| Patch images | McNeal's anatomy study[32] (15 cases), ~512x512 µm | 11,862+ | Philips | 10x |
| *Recent slides (<6 years)* | | | | |
| TMA Spots (Smallest sample dimension) | Stanford TMA database (339 cases), 2048x2048 pixels spot images | 1,129 | Leica | 20x |
| All whole mount slides of each single case+++ (Largest sample dimension) | Radical prostatectomy (46 cases), whole-mount slide | 368 | Leica | 20x |
| **Tumor volume estimation** | | | | |
| All slides of each single case+++ | Radical prostatectomy (46 cases), whole-mount slides | 368 | Leica | 20x |
| **Sorting slides according to prostate cancer presence** | | | | |
| *Tissue sampling method* | | | | |
| Radical prostatectomy specimens | Radical prostatectomy (136), whole-mount slides | 1,080 | Leica | 20x |
| Dissected pelvic lymph nodes | Lymph node dissection (50 lymph nodes), whole slides | 19 | Leica | 20x |
| Obduction Cystprostatectomy specimens | The Genotype-Tissue Expression project [33] (40 cases), whole slides++ | 40 | Leica | 40x |
| Radical prostatectomy specimens | PAI-WSIT project [34] (18 cases), whole slides++ | 60 | Hamamatsu | 40x |
| **Gleason pattern detection and ISUP grading** | | | | |
| *Tissue sampling method* | | | | |
| Biopsy cores | The International Society of Urological Pathology image library [35], 2,048x2,048 pixels (72dpi) micrograph, (137 cases) | 137 | Olympus (micrograph) | 20x |
| Radical prostatectomy specimens | Radical prostatectomy (136), whole-mount slides | 1,080 | Leica | 20x |
| Very limited tissues with prostate cancer | Patches from 594 random regions with Gleason patterns 3-5 and HGPIN in 24 whole-mount WS images (24 cases), 512×512 $\mu$m | 3,840 (20x)<br>1,128 (10x) | Leica | 20x<br>10x |
| **Detection of ductal morphology** | | | | |
| Patch images | Patches from 38 random regions from 2 WS images of 2 cases with ductal adenocarcinoma plus 218 random regions with Gleason pattern 3-5 in 9 WS images of 9 cases, 512x512 $\mu$m | 2,112 | Leica | 10x |
| **Detection of cribriform pattern** | | | | |
| Patch images | Patches from 32 random regions with Cribriform patterns of 5 cases and 199 random regions with non-cribriform prostate cancers and Gleason patterns 3-5 in 9 whole-mount WS images (9 cases), ~512×512 $\mu$m | 928 | Leica | 10x |
| **Detection of vessels** | | | | |
| Patch images | Patches from 642 random regions with blood vessels on 22 WS images (22 cases) and 478 random regions with Gleason patterns 3-5 on 20 WS images (20 cases), 512x512 $\mu$m | 4,608 | Leica | 10x |
| **Detection of nerve structure** | | | | |
| Patch images | Patches from 628 random regions with nerves or ganglions on 22 WS images (22 cases) and 216 random regions with Gleason patterns 4-5 (8 slides, 8 cases), 512x512 $\mu$m | 1,280 | Leica | 10x |
| **Detection of inflammatory cell infiltration** | | | | |
| Patch images | Patches from 123 random regions with inflammatory cell infiltration on 19 WS images (19 cases) and 216 random regions with Gleason patterns 4 and 5 (8 slides, 8 cases), ~512x512 $\mu$m | 768 | Leica | 10x |
| **HGPIN detection** | | | | |
| Patch images | Random 32 regions from 10 WM images (10 cases); 40 random regions with intraductal adenocarcinoma from 4 WM images (4 cases); 19 random regions with benign prostatic hyperplasia from 4 WM images (4 cases), ~512x512 $\mu$m | 2,687 | Leica | 10x |
| **Integration into an electronic pathology report platform** | | | | |
| Radical prostatectomy specimens | 136 radical prostatectomy specimens, complete representative whole-mount slides per case | 1,028 (Median: 8 per case) | Leica | 20x |

## Image processing

Whole slide images are gigapixel images and difficult to process in one step. Therefore, we tiled the whole slide images into small patches after identifying the foreground prostatic tissue using image thresholding according to Otsu's method[36]. We obtained a final patch size of 512×512 pixels at 10× by a pixel mapping of 1 $\mu$m per pixel. For Gleason pattern detection, the 20× objective magnification (~256×256 $\mu$m) was also evaluated to identify the optimal magnification level for the agreement between vPatho and the pathologist.

A patch was labelled as positive for PCa when more than one percent of the patch was positive. A patch was positive for high-grade prostatic intraepithelial neoplasia (HGPIN), Gleason patterns, ductal morphology, cribriform pattern, nerves, or vessels when at least 5% of the patch was positive.

Spatial annotation data curated by pathologists were used to label these patches. For patch-wise evaluation, each histology image has a corresponding mask that incorporates the demarcated lesion areas and has the dimension equal to the dimension of the original image. Both the image and the mask were tiled using the same grid.

In contrast, the spot images from TMA were first downsized to achieve 10x magnification level; then, prostatic tissue was identified by applying image thresholding according to Otsu's method and boundary was determined using the contour detection algorithm provided by OpenCV framework[37]. After that, the region of interest was divided into tiles by 512×512 pixels (512×512 $\mu$m).

Each TMA spot (each spot was captured at 20x objective magnification) was previously labelled for PCa presence based on the pathologist's judgment and whether it originated from cancer lesions in prostatectomy specimens. Accordingly, each patch was labelled based on the spot label.

### Quality assessment

The quality assessment of slides by technical assistants and pathologists is an integral component of the standard operating procedure of accredited pathology institutions[38]. In clinical settings, when a slide is not suitable for pathology evaluation, a better slide is prepared from the same tissue block (prostatectomy specimens are embedded in blocks) for pathology evaluation. We anticipate that similar standards for image quality assessment will become routine in future digital pathology workflows. However, the implementation of a quality management system for histology slides in clinical setting is beyond the scope of the current study. We consequently refer the reader to the relevant literature.

Although all histology images passed internal review prior to inclusion in the current study, we implemented a blurriness and illuminance assessment tool for patches.

The blurriness of each patch was estimated using the variance of the Laplacian[39]; we first established a reference range (95% Confidence interval: 112 - 124) for blurriness detection on 800 patches randomly generated at 10× objective magnification from 8 diagnostic H&E WS images available in TCGA-PRAD (100 patches extracted from random regions for each WS image). These slides did not contain visible blurriness or illuminance imbalance to human eyes. If a patch had a variance outside this range, image sharpening[40] was applied to reduce the blurriness of the patch. A second blurriness assessment was then made, and if the variance of the patch was still outside the reference range, the patch was excluded from the detection task.

In parallel, we assessed the relative illuminance[41] of each patch and corrected its illuminance if the relative illuminance was outside the reference range (144.4 – 165.3). We determined the reference range on 800 patches previously utilized to identify the reference range for blurriness. For illuminance correction, we applied automated Contrast Limited Adaptive Histogram Equalization (CLAHE)[42] in combination with a modified version of a reference-free Macenko approach for stain color optimization [43]. Our aim with the illuminance correction was to reduce the illuminance deviation from the reference range. Finally, we evaluated the impact of the correction of blurriness and illumination on the detection performance for spots with prostate cancer using Stanford's tissue microarray with 1,129 spots.

## Model Architecture

We utilized a novel convolutional neural network called PlexusNet[44] that supports neural architecture search (NAS)[45] and uses ResNet[46] and Inception blocks[47] as well as standard convolutional blocks (VGG). Taking inspiration from a previous study, we also included quasi soft attention blocks[48] (we remove "quasi" in the block description for convenience). The PlexusNet architecture is a directed acyclic multigraph. The complexity of this graph is primarily determined by its depth (i.e., the number of levels), the extent of its branching (i.e., end-to-end paths or width) as well as the number of weighted junctions (i.e., cross edges) between two end-to-end paths that give arise to the multigraph property. A transitory "short" path was randomly defined to populate feature maps of an end-to-end path at a random level (**Figure 6**). The resulting feature maps (i.e., channels) for all end-to-end paths were concatenated and their feature dimensions were then reduced using global pooling (either maximum or average pooling). The classification section of PlexusNet fully connects and weights the dense features to feed into the final layer to estimate the confidence scores for pathology existence in patches of histology images. **Figure 6** provides a summary of the PlexusNet architecture concept and **Figures 7 and 8** illustrate different block types used in the PlexusNet architecture. **Table 3** summarizes the hyperparameter configurations for each finding.

We used Equation 1 to justify the contrast and interpolation of the input images as this step improved the classification performance for prostate cancer detection by 10.0% (95% CI: 9.2 – 11.2%) compared to a convnet model without this interpolation function.

**Equation 1.**

$$\hat{X} = -2e^{-(2X^2)}[\cos(90\ \omega_1) + x\ \sin(90\ \omega_2)]$$

$\hat{X}$ is the output of the equation, where $\omega_1$ and $\omega_2$ are the trained weights (scalar) and $\omega_1, \omega_2 \in [-1,1]$ and X is a matrix represents an image batch defined by $\mathrm{X} = \{\mathrm{x} \mid \forall x \in X, 0 \leq x \leq 1\}$; $\cos(90\ \omega_1)$ and $\sin(90\ \omega_2)$ terms are the interpolation functions. The optimal weights $\omega_1$ and $\omega_2$ are determined during model training.

**Equation 2.**

$$\tilde{X} = 2\frac{\hat{X} - \min(\hat{X})}{\max(\hat{X}) - \min(\hat{X})} - 1$$

This equation does a [-1,1] feature scaling where $\tilde{X}$ is the normalized output and $\hat{X}$ is the batch input that is interpolated by Eq. 1. **Figure 9** illustrates the results of Eq. 1 on the input image.

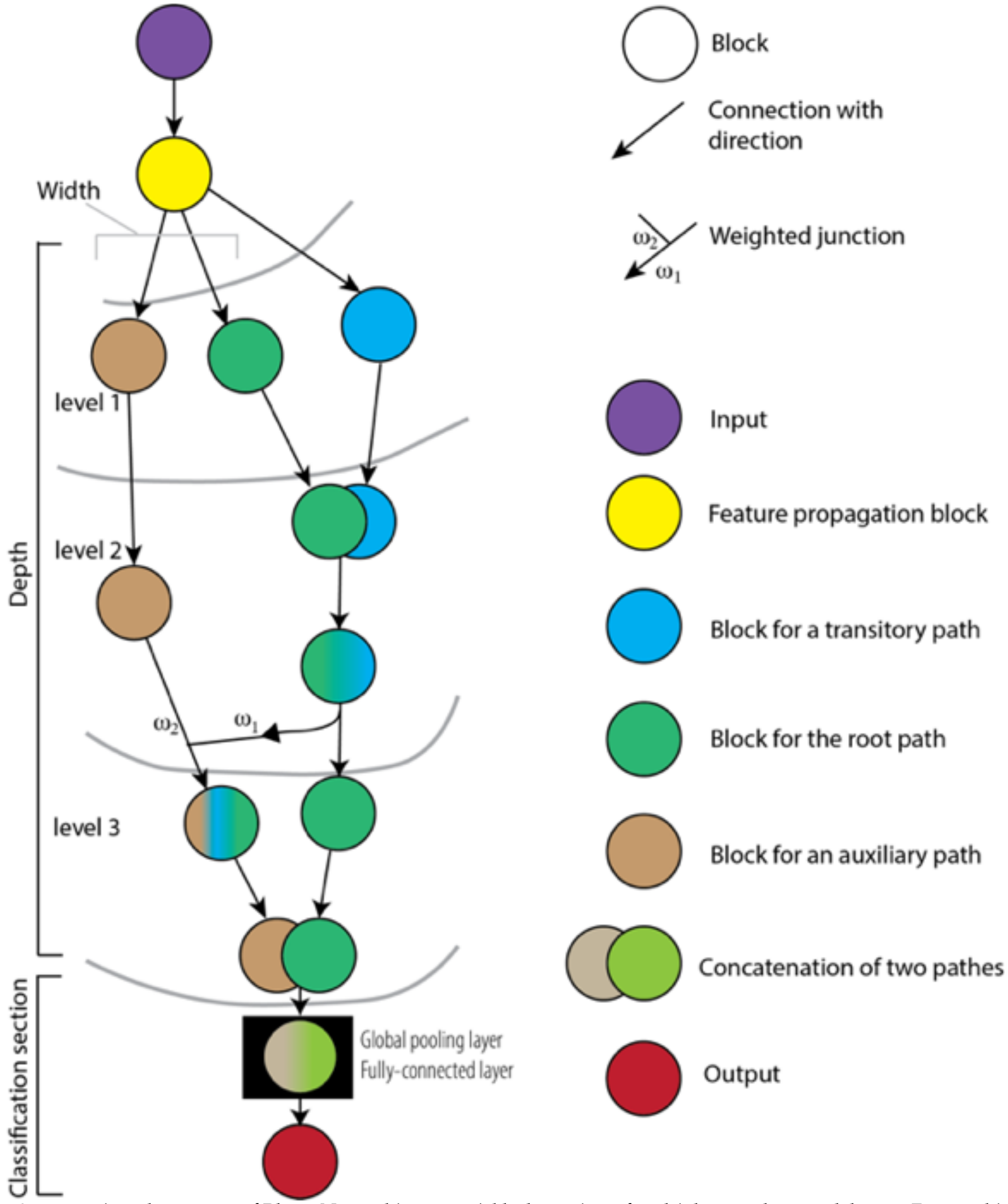

*Figure 6 summarizes the concept of PlexusNet architecture. A block consists of multiple neural network layers. Four architecture block types are available: VGG, Inception, residual, or soft attention block. The major hyperparameters for the graph definition are the depth (number of levels), with a minimum depth of 2, the number of end-to-end paths (width), the number of transitory "short" paths, and junctions that intersect between two end-to-end paths. Here, the example PlexusNet architecture has a depth of 3 levels, its width is 2, and it has a single transitory path and weighted junction between two end-to-end paths. The position of the weighted junction between two paths before the global pooling layer is determined randomly. For all PlexusNet models, all final feature maps of end-to-end paths are concatenated before feeding into the Global pooling layer. The depth of a transitory path is determined randomly, and the transitory path concatenates with the root path (By default, the first end-to-end path is considered as root path) at the same level of the weighted junction. The position randomization for weighted junctions or the depth for transitory path has no impact on the model performance while the number of weighted junctions has impact on the model performance. For a simplified model development, we unified the block type for all paths.*

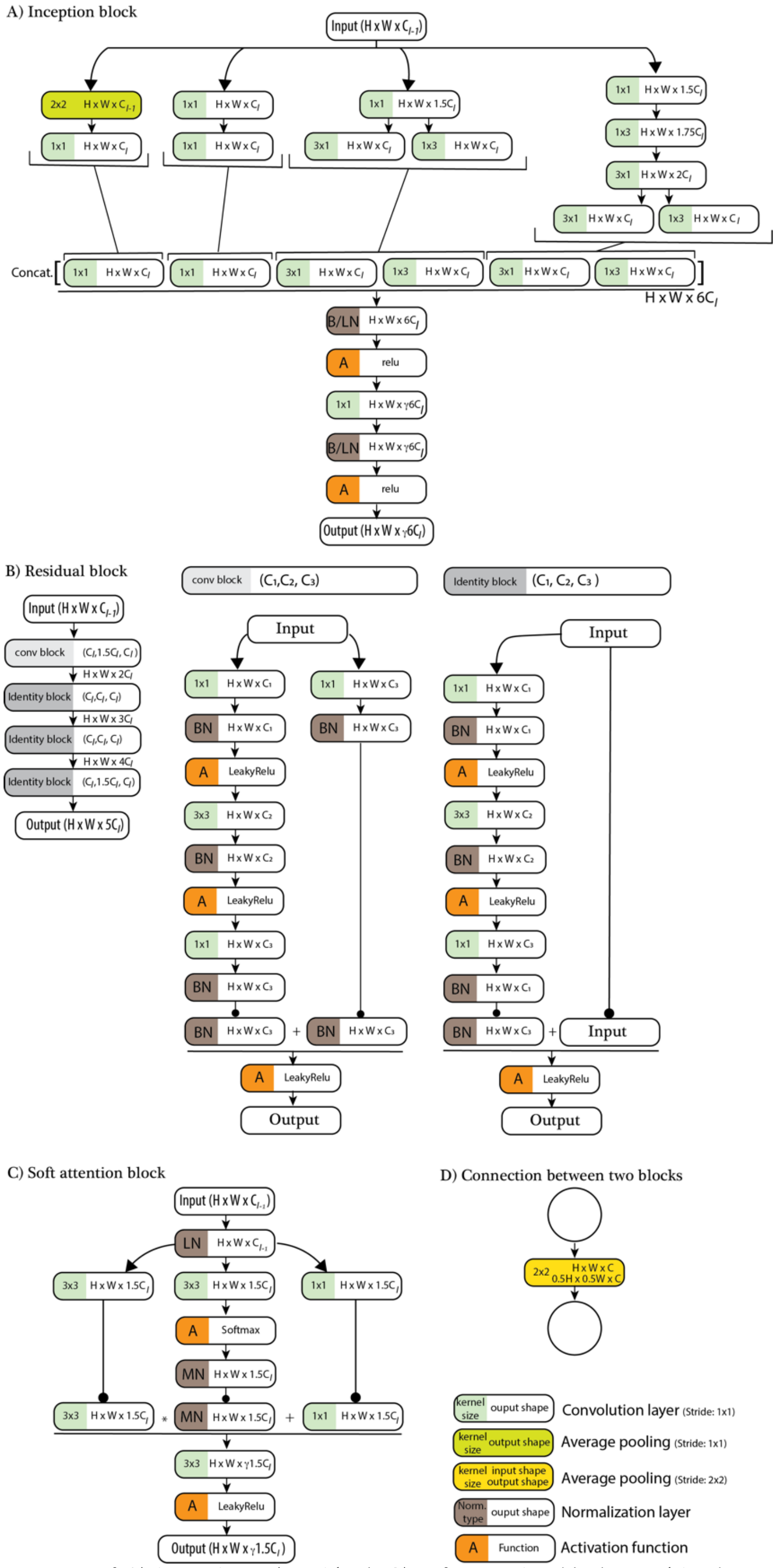

*Figure 7 illustrates the structures of A) inception, B) residual, C) soft attention blocks used in the current study. In PlexusNet architecture, (D) two consecutive blocks are connected by an average pooling that reduces the width and heights of the feature maps of the next block by half. When we calculate the channel numbers, the "round half up" approach is used to convert them to an integer number. BN: Batch normalization; MN: max normalization; LN: Layer normalization; B/LN: either batch or layer normalization. $\gamma$ is the compression rate to reduce the channel information similar to the compression ratio in DenseNet [49]; l: the level index. H: Height; W: width; C: Channel. The subscript of the level definition for H and W was ignored to emphasize that H and W did not change during the tensor processing in each block.*

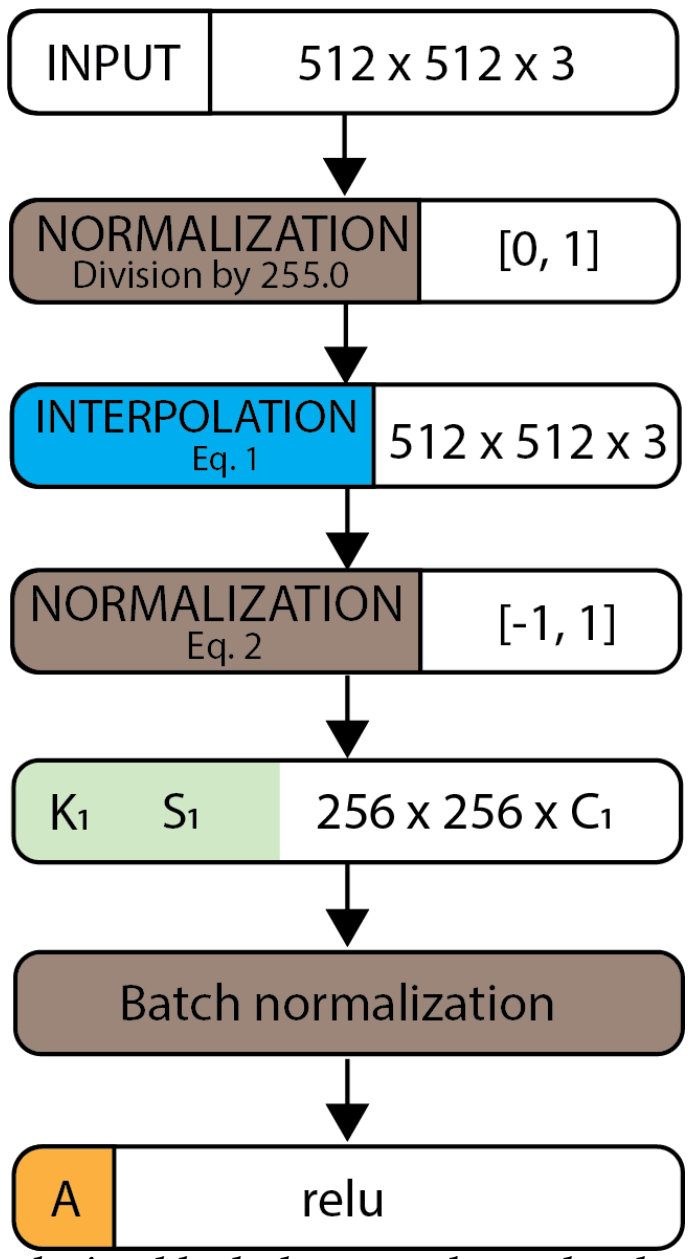

*Figure 8 represent the layers for the feature population block that populates the channel number from 3 to $C_1$ using the convolution layer with the size of the convolution weight kernel ($K_1$) and a Stride ($S_1$). The default value for stride $S_1$ is 2×2 pixels to move $K_1$. The dimension unit is given in pixel.*

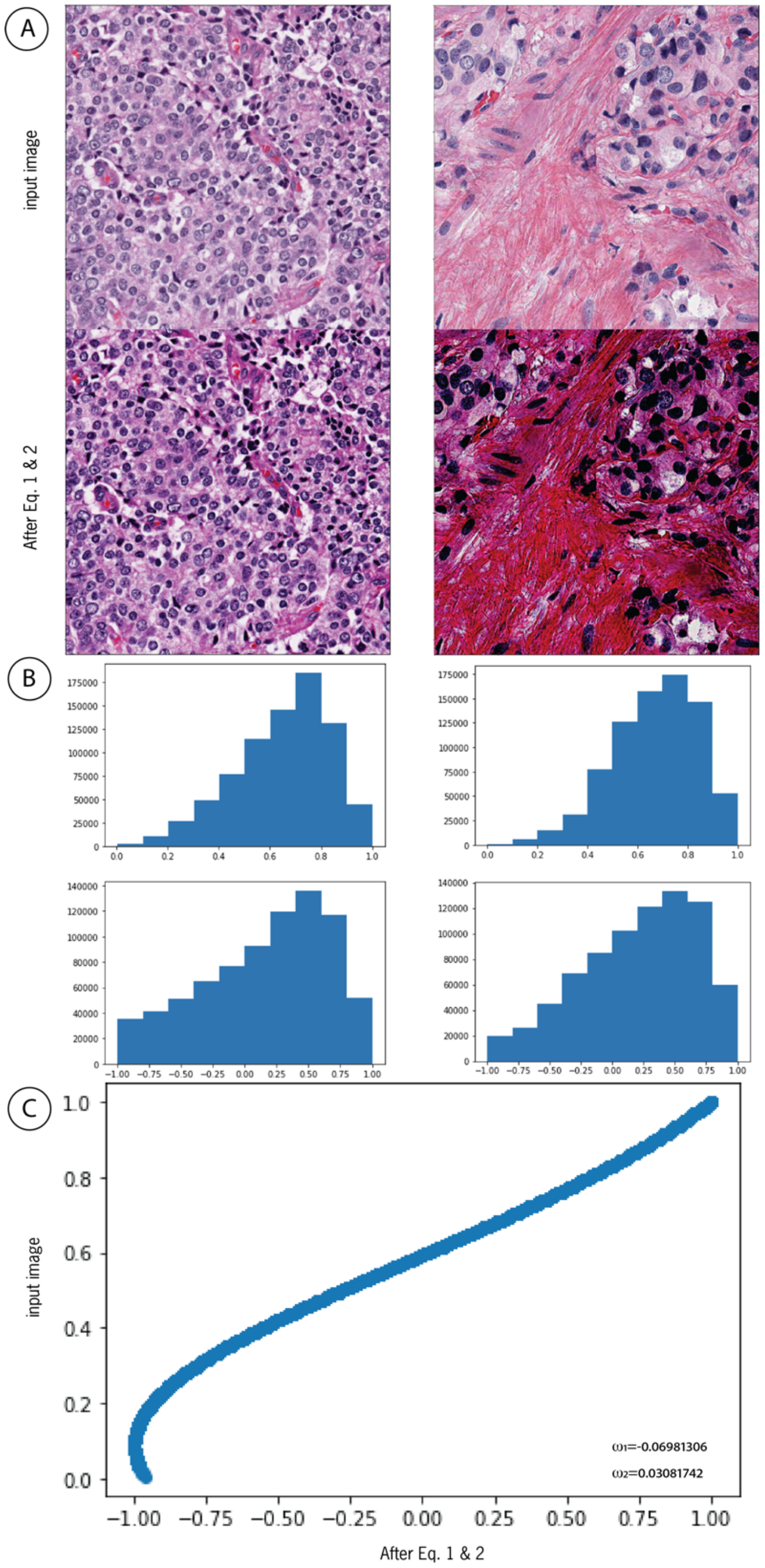

*Figure 9: (A) illustrates the same image before and after applying Equation (Eq.) 1 and 2 using $\omega_1, \omega_2$ provided in (C) for comparison. (B) histograms of the same image before and after applying Eq. 1 and 2. The application of Eq. 1 remarkably improves the image contrast and (B) interpolates the input image according to $\omega_1, \omega_2$. PlexusNet-based models learn the optimal values $\omega_1, \omega_2$ during the model training to determine the optimal non-linear interpolation of the input image to solve a classification problem. The intuition behind this is to contrast the semantic content of the input image to increase the likelihood of capturing meaningful features in the deep convolutional neural network layers that consequently impact the classification performance; $\omega_1, \omega_2$ given in (C) were originated from the prostate cancer detection model. The example patch image has a dimension of ~512 µm × 512 µm at 10× objective magnification. (C) values of the input image were normalized using min-max normalization and correlated with the values from Eq. 1 and 2.*

*Table 3 summarizes the architecture design for different findings. By combining all the models listed here, the novel architecture design has achieved a total parameter capacity remarkably lower than the parameter capacities of a single ResNet-18 model (~11 millions) or the 2nd version of a single MobileNet model (~2.0 millions parameters) [50]. These models have combined a total of 1,012 features in the fully-connected layers (In comparsion, a single RestNet-18 model has 512 features in the fully-connected layers). Given the compactness of our models, we could assign all 17 models to a single GPU card (VRAM 24 GB) for pathology report generation task. For each Gleason pattern, we applied an ensemble modeling that weighted the predictions of the models equally. The weighted prediction scores ranged between 0 and 1. Note: The variation in the parameter capcity despite having the same hyperparameter configuration is due to the variation in the block number (depth) of the short path. Approximately a hunderd of models with different configuration were experimented to achieve the final model configuration using the grid search and trial-and-error approach.*

*Ch.: Channel; conv.: Convolution; $C_1$: Channel number of the first convolution layer; $K_1$: the kernel size of the first convolution layer; act.: activation; no.: number; cat. : category; FC: fully connected; +: Layer normalization[51] is applied instead of batch normalization[52]. Transfomer blocks were added prior the Global pooling; since the output of the convolutional layer is Batch size × Height × Width × Channel, we reshaped the output to Batch size × Height ∗ Width × Channel before feeding into the Transformer block; the output from Transformer[48] was reshaped back to the dimension of the convolutional layer before a global pooling was applied. For all models, a patch dimension of 512 × 512 × 3 is applied.*

| Architecture / Finding | *Block type* | *Depth* | *Width* | *Junction* | *Short path* | $C_1$ | $K_1$ | *Apply crop center on input (256×256)* | *Initial filter factor per path* | *Global pooling* | No. of Ch. for 1st FC layer | Output function (no. of cat.) | *Parameter capacity* | *Supervised contrastive learning[53]* | *Transformer[48] (no. of blocks. header number)* |
|---|---|---|---|---|---|---|---|---|---|---|---|---|---|---|---|
| Prostate Cancer | Inception | 7 | 2 | 3 | 1 | 32 | 5 × 5 | No | 2 | Max. | 96 | Softmax (2) | 178,342 | No | No |
| Gleason pattern 3 | Inception | 4 | 2 | 3 | 1 | 32 | 5 × 5 | No | 2 | Max. | 60 | Softmax (2) | 58,879 | No | No |
| | Inception | 4 | 2 | 3 | 1 | 32 | 5 × 5 | No | 2 | Max. | 60 | Softmax (2) | 58,879 | No | No |
| Gleason pattern 4 | Inception+ | 4 | 2 | 3 | 1 | 6 | 5 × 5 | No | 2 | Avg. | 60 | Sigmoid (1) | 49,122 | Yes | No |
| | Inception+ | 4 | 2 | 3 | 1 | 6 | 5 × 5 | No | 2 | Avg. | 60 | Sigmoid (1) | 49,122 | Yes | No |
| | Inception+ | 4 | 2 | 3 | 1 | 32 | 5 × 5 | No | 2 | Avg. | 60 | Softmax (2) | 50,201 | No | No |
| | Soft attention | 6 | 2 | 3 | 1 | 8 | 3 × 3 | Yes | 2 | Max | 20 | Softmax (2) | 51,732 | No | No |
| | Soft attention | 6 | 2 | 3 | 1 | 8 | 3 × 3 | Yes | 2 | Max | 20 | Softmax (2) | 51,732 | No | No |
| | ResNet | 5 | 2 | 3 | 1 | 16 | 5 × 5 | No | 4 | Avg. | 44 | Softmax (2) | 206,638 | No | No |
| Gleason pattern 5 | Soft attention | 5 | 2 | 3 | 1 | 16 | 5 × 5 | No | 2 | Avg. | 24 | Softmax (2) | 48,590 | No | No |
| | Soft attention | 5 | 2 | 3 | 1 | 16 | 5 × 5 | No | 2 | Avg. | 24 | Softmax (2) | 48,590 | No | No |
| *Ductal morphology* | Inception | 4 | 2 | 3 | 1 | 16 | 5 × 5 | No | 2 | Max | 60 | Softmax (2) | 53,683 | No | No |
| *Cribriform pattern* | Inception | 3 | 2 | 3 | 1 | 32 | 5 × 5 | No | 2 | Max | 48 | Softmax (2) | 34,033 | No | No |
| HGPIN | Inception | 5 | 4 | 2 | 2 | 16 | 5 × 5 | No | 4 | Max | 160 | Softmax (2) | 412,322 | No | No |
| Vessel | Inception+ | 5 | 2 | 3 | 1 | 16 | 5 × 5 | No | 2 | Max | 72 | Softmax (2) | 183,448 | No | Yes (3,4) |
| Nerve | Inception+ | 5 | 2 | 3 | 1 | 16 | 5 × 5 | No | 2 | Max | 72 | Softmax (2) | 171,997 | No | Yes (3,4) |
| Inflammatory cell infiltration | Inception+ | 5 | 2 | 3 | 1 | 16 | 5 × 5 | No | 2 | Max | 72 | Softmax (2) | 171,277 | No | Yes (3,4) |
| ***Total*** | | | | | | | | | | | **1,012** | | **1,878,587** | | |

We considered PlexusNet architecture as this architecture facilitates developing small models (See parameter capacity in **Table 3**) for accurate binary classification tasks comparable to the large state-of-art models as shown in **Supplementary file 1**. It is worth mentioning that the cumulative parameter capacity of the models we considered for vPatho is remarkably below the parameter capacity of a single ResNet 18 model (18 million trainable parameters), a frequently used model architecture. Moreover, the parallel use of these models is feasible on a single GPU card with 24 GB, when we have a batch size of 16 patches (Dimension: 16×512×512x3).

## Datasets for model development

For training, we followed a data-efficient strategy where we pre-defined and determined the proportion of the pathological content of the training set. We followed a trial-and-error approach to determine the optimal proportion of the pathological content of the training set for each detection task. We mitigated the imbalanced classification conditions of the training set by oversampling the underrepresented positive finding and applied image augmentation with 50% probability to vary the content visualization of the patches. **Tables 4 to 10** summarize the data compositions for model development.

For precursor detection, cancer morphology detection and mesenchymal structure detection, we intentionally defined Gleason pattern 5 and benign prostatic hyperplasia as unseen pathological findings and included patches of these findings in the optimization data sets. The aim of this strategy was to increase the likelihood of detecting best and well-fitted models.

*Table 4: the non-overlapping patch number and the cancer proportion resulting from splitting 200 whole-slide images (WSI) at 10x objective magnification by 512 × 512 pixels (1 pixel corresponds to ~1 µm) obtained from the Cancer Genome Atlas image library for prostate cancer (PRAD). All slides were scanned at 40x objective magnification. We randomly selected 200 WSI for model development where we considered 195 WSI for training and 5 WSI for in-training optimization. These numbers were fixed when curating different folds by random data splitting). All images were stained with H&E.*

| **Finding** | **Random data splitting** | | |
|---|---|---|---|
| | Fold 1 | Fold 2 | Fold 3 |
| **Training set (case number=195)** | | | |
| Patches | | | |
| *Non-cancer tissues, n (%)* | 44,240 (53.99) | 43,530 (53.35) | 43,975 (54.03) |
| *Prostate Cancer, n (%)* | 37,69 (46.01) | 38,069 (46.65) | 37,411 (45.97) |
| *Total, n (%)* | 81,937 (100.00) | 81,599 (100.00) | 81,386 (100.00) |
| Average cancer pixel proportion in a patch labelled with prostate cancer (median) | 78.8% (100.0%) | 78.9% (100.0%) | 78.9 % (100.0%) |
| **In-training optimization set (case number=5)** | | | |
| Patches | | | |
| *Non-cancer tissues, n (%)* | 1,769 (72.0) | 2,479 (88.7) | 2,034 (67.6) |
| *Prostate Cancer, n (%)* | 689 (28.0) | 317 (11.3) | 975 (32.4) |
| *Total, n (%)* | 2,458 (100.0) | 2,796 (100.0) | 3,009 (100.0) |
| Average pixelwise cancer proportion in a patch labelled with prostate cancer (median) | 76.7% (100.0%) | 64.9% (72.0%) | 75.7 % (97.0%) |

*Table 5: The patch number and the cancer proportions in the external datasets curated from 60 whole-mount histology images. Please be aware that one whole-mount image covers the whole prostatic slice and provides patches on average 6 times more than a single whole-slide image of the Cancer Genome Atlas for prostate cancer.*

| **External data set for model comparison (60 whole mount images)** | |
|---|---|
| **Finding** | **Patches** |
| *Non-cancer tissues, n (%)* | 142,135 (81.95) |
| *Prostate Cancer, n (%)* | 31,286 (18.05) |
| *Total, n (%)* | 173,421 (100.00) |
| Average pixelwise cancer proportion in a patch labelled with prostate cancer (median) | 80.6% (100%) |

*Table 6: the number of patches generated from 641 tissue microarray spot images (2240x2240 pixels, scanned at 20x objective magnification) provided by Arvaniti et al with spatial annotation for prostate cancer and Gleason patterns prepared by two pathologists[6]. The overlap between two patches was predefined to be 50%.*

| Finding | Magnification level | Patch number, n (%) |
|---|---|---|
| Non-cancerous tissues | 5x | 532 (35.9) |
| | 10x | 3325 (35.9) |
| | 20x | 4166 (35.9) |
| Gleason pattern 3 | 5x | 380 (25.7) |
| | 10x | 2375 (25.7) |
| | 20x | 2932 (25.7) |
| Gleason pattern 4 | 5x | 328 (22.2) |
| | 10x | 2049 (22.2) |
| | 20x | 2867 (22.2) |
| Gleason pattern 5 | 5x | 240 (16.2) |
| | 10x | 1500 (16.2) |
| | 20x | 1732 (16.2) |

*Table 7: The consitutions of training sets on the basis of Table 4 aims to cover the variation in the magnification levels for the model development. Various approaches for patch augmentation were applied to increase the variation in the patch appearance to increase the likelihhod for more generalizable models.*

| Finding/Models | Considered datasets with magnification levels as training set | Augmentation features (Augmentation probability: 50%) |
|---|---|---|
| GP3 | | Random brightness and contrast<br>Random image compression rates (variable image resolution)<br>Random flip (horizontal and vertical)<br>Random rotation (between -90 and 90)<br>Random hue saturation value<br>Random Gaussian noise<br>Random clip limits in the Contrast limited adaptive histogram equalization[54] |
| Model 1 | 5x, 10x, 20x | |
| Model 2 | 10x ,20x | |
| GP4 | | |
| Model 1 | 10x, 20x | |
| Model 2 | | |
| Model 3 | | |
| Model 4 | | |
| Model 5 | | |
| Model 6 | | |
| GP5 | | |
| Model 1 | 10x | |
| Model 2 | | |

*Table 8: The optimization set for Gleason patterns detection used to select the best models according to the best AUROC. This optimization set was curated from The Cancer Genome Altas images with demarcation of 536 prostate cancer hetergenous regions representing different Gleason patterns and benign tissues in 35 cases (~15 regions per one case) by a team of a senior pathologist and a prostate cancer researcher. These patches (512x512 pixels) were curated at 10x magnification level, resulting in a total of 8308 patches. The categorcy "Normal tissue" covers prostatic epithelium, stroma, and atrophy. HGPIN stands for high-grade prostatic intraepithelial neoplasia is the presumed precursors of prostate cancers.*

| Finding | Case, n (%) | Regions, n (%) | Patches, n (%) |
|---|---|---|---|
| GP3 | 12 (21.4) | 32 (6.0) | 923 (11.1) |
| GP4 | 9 (16.1) | 166 (31.0) | 1318 (15.9) |
| GP5 | 20 (35.8) | 216 (40.2) | 5397 (65.0) |
| Normal tissue | 11 (19.6) | 106 (19.8) | 466 (5.6) |
| HGPIN | 4 (7.1) | 16 (3.0) | 204 (2.4) |

*Table 9: The training sets used to develop patch-level classification models for cribriform pattern, ductal morphology, high-grade intraprostatic intraepithelial neoplasia (HGPIN), vessels, nerve, inflammatory cell infiltration. A total of 42 cases and 1,723 regions were considered. For all findings, we oversampled the positive patches to encounter the class imbalance. + 10 additional images were collected from internet as ductal adenocarcinoma is a rare type of prostate cancer. Given that benign prostatic hyperplasia may have glandular structure with large lumen, we explicitly excluded patches from the lumen. Gleason patterns and normal glandular tissues further included stromal components. Normal tissues cover ductus deference, epithelial and stromal components from peripheral and central zones of prostate. Gleason pattern 5 was not included in the training set as this pathology was used as unseen finding in optimization set to identify ideal model. The initial HGPIN model was retrained on a training set that had additionally 4,143 patches with prostatic hyperplasia from 19 lesions of 4 cases. Here, we used a learning rate of 1e-6 to avoid a complete distortion of the initial weights[55].*

| Finding | Patches | | | | |
|---|---|---|---|---|---|
| | regions, n (case, n) | Magnification level | Positive, n (%) | Negative, n (%), (regions, n; cases, n) | List of negative findings (%) |
| Cribriform pattern | 117 (13) | 10x | 546 (24.5) | 1686 (75.5) (672;29) | -Gleason pattern 3 (45.0)<br>-Nerves (35.3)<br>-Normal glandular tissues (11.4)<br>-Gleason pattern 4 No Cribriform pattern (8.3) |
| Ductal morphology | 83 (19) + | 10x | 278 (12.7) | 1795 (87.3) (218; 21) | -Gleason pattern 3 (51.4)<br>-Gleason pattern 4 (48.6) |
| HGPIN++ | 40 (4) | 10x | 248 (3.0) | 5009 (97) (1564,37) | -Gleason pattern 3 (9.2)<br>-Gleason pattern 4 (13.3)<br>-Cribriform pattern (11.6)<br>-Nerves (16.1)<br>-Vessel (12.7)<br>-Ductal adenocarcinoma. (8.9)<br>-Normal tissues (5.4)<br>-Inflammatory cell infiltration (22.3) |
| | | | | | -Perineural invasion.<br>(0.5) |
| Blood vessels | 232 (23) | 10x | 375 (10.3) | 3280 (89.7)<br>(955;42) | -Gleason pattern 4<br>(28.2)<br>-Gleason pattern 3<br>(22.8)<br>-Nerves<br>(18.1)<br>-Cribriform pattern<br>(16.8)<br>-Ductal adenocarcinoma<br>(8.0)<br>-Normal glandular tissues<br>(6.1) |
| Nerves | 344 (20) | 10x | 389 (11.2) | 3088 (88.8)<br>(545;31) | -Gleason pattern 4<br>(28.9)<br>-Gleason pattern 3<br>(23.9)<br>-Cribriform pattern<br>(18.5)<br>-Blood vessels<br>(14.6)<br>-Ductal adenocarcinoma<br>(8.0)<br>-Normal glandular tissues<br>(6.0) |
| Inflammatory cell infiltration | 418 (25) | 10x | 430 (10.5) | 3663 (89.5)<br>(867;35) | -Gleason pattern 4<br>(24.7)<br>-Gleason pattern 3<br>(20.0)<br>-Nerves<br>(16.5)<br>-Cribriform pattern<br>(15.5)<br>-Blood vessels<br>(12.6)<br>-Ductal adenocarcinoma<br>(7.0)<br>-Normal glandular tissues<br>(3.6) |

*Table 10: The optimization sets used to select the best models for high-grade intraprostatic intraepithelial neoplasia, vessels, nerve, and inflammatory cell infiltration. These dataset was curated from the PRAD-TCGA histology image dataset. We emphasize that we added images representing Gleason pattern 5 as unseen pathological finding. A total of additional 42 cases and were considered for optimization set as well. For HGPIN, we increased the patch number to increase*

| Finding | Regions, n (Case, n) | Magnification level | Positive, n (%) | Negative, n (regions, n; cases, n) | List of negative findings (%) |
|---|---|---|---|---|---|
| Cribriform pattern | 87 (12) | 10x | 197 (32.9) | 402 (67.1) (197,25) | -Gleason pattern 3 (41.0)<br>-Gleason pattern 4 without cribriform pattern (9.4)<br>-Nerve (36.7)<br>-Normal glandular tissues (12.9) |
| Ductal morphology | 36 (1) | 10x | 59 (3%) | 1,906 (97%) (496, 42) | -Gleason pattern 3 (8.3)<br>-Gleason pattern 4 (11.4)<br>-Gleason pattern 5 (56.1)<br>-Nerve (8.6)<br>-Blood vessel (6.8)<br>-Cribriform pattern (6.9)<br>-Normal glandular tissues (1.9) |
| HGPIN | 16 (7) | 10x | 204 (2.5) | 8,104 (97.5) (581, 35) | -Benign Prostatic Hyperplasia (5.8)<br>- Gleason pattern 3 (11.4)<br>-Gleason pattern 4 (16.2)<br>-Gleason pattern 5 (66.6) |
| Blood vessel | 96 (18) | 10x | 123 (6.3) | 1,820 (93.7) (447,40) | -Gleason pattern 3 (9.6)<br>-Gleason pattern 4 (10.9)<br>-Gleason pattern 5 (57.8)<br>-Nerve (8.2)<br>-Cribriform pattern (7.9)<br>-Normal glandular tissues (2.2)<br>-Ductal adenocarcinoma (3.4) |
| Nerve | 147 (17) | 10x | 157 (8.0) | 1,812 (92.0) (392,41) | Gleason pattern 3 (10.2)<br>Gleason pattern 4 (12.7)<br>Gleason pattern 5 (56.8)<br>Cribriform pattern (6.7)<br>Vessel (6.3)<br>Normal glandular tissues (3.1)<br>Ductal adenocarcinoma (4.2) |
| Inflammatory cell infiltration | 109 (22) | 10x | 161 (7.8) | 1904 (92.2) (511, 42) | Gleason pattern 3 (10.0)<br>Gleason pattern 4 (11.4)<br>Gleason pattern 5 (54.1)<br>Nerve. (7.3)<br>Cribriform pattern (6.5)<br>Vessel (5.5)<br>Normal glandular tissues (1.7)<br>Ductal adenocarcinoma (3.5) |

## Comparison with state-of-the-art model architecture

As exploratory analysis to determine the optimal model architecture and to justify our model architecture selection, a comparison of our novel model architecture with the state-of-the-art model architectures was conducted on 60 randomly selected whole mount images from Stanford to detect tiles with prostate cancers (a patch is 512×512 pixels and corresponds to ~512×512 $\mu$m at 10× objective magnification). We compared the accuracy performance on non-overlapping patches generated from these images as patches represent the smallest data unit on which different detection tasks depend. Models were trained using categorical cross entropy loss with a batch size of 16 and optimized using "ADAM" with a default configuration[56] and learning rate of 1e-3. The patch augmentation incorporated random rotation, JPEG compression rates for random image resolution, flipping and color shifting as well as zooming. For all procedures, we utilized the same random seed (seed=1234) to ensure that patch augmentation was similarly applied for all models. To account for variability in development set portioning, we partitioned the development set in 3 folds and repeated our evaluation three-time on held-out test set. In every fold, we evaluated the area under the receiver operating characteristic (AUROC), expected classification error, Brier score for patches per slide and then determined the mean and 95% Confidence interval using 100,000 bootstrapped slide resampling. The expected classification errors (ECE) and Brier scores provide insights into the model goodness of fit and the model calibration. The lower the ECE or Brier score, the better the goodness of fit and calibration of the model. From our evaluation, we identified that our novel model "PlexusNet" achieved comparable performance to the state-of-art models widely used in medical imaging while its parameter capacity was 150- or 85-times smaller than ResNet-50v2 or VGG16; the per-batch training duration for PlexusNet was also at least two-times shorter (223±54 milliseconds per batch vs. 630±67 milliseconds per batch for RestNet-50v or 528±68 for VGG-16) on a single GPU card (NVIDIA™ Titan Volta with 11 GB) under similar data input/output condition (one training process at a time and running processes relevant for the operating system). The results of comparison analysis are summarized in **Supplementary File 1**.

## Hyperparameter configuration for model training

Table 11 provides the hyperparameter information used during model training. Models with the lowest AUROC on the optimization set were considered. For supervised contrastive learning, we selected the model with the lowest loss value in the steps for contrastive feature learning[53,57].

*Table 11 shows the final hyperparameter configuration determined according to trial-and-errors approach and grid search that resulted in examining 100 models. ** the epoch of models that we considered as final models during the development setting. AUROC was determined on the optimization set. Models were selected based on their AUROC. $^{+}$ highlights that we consider two models originated from the same training setting. The reasons of considering multiple models are described in the following section.*

| Model | loss function | Batch size | Optimization algorithm | Best epoch** (Max epochs) | Initial learning rate | AUROC * |
|---|---|---|---|---|---|---|
| *GP3* | | | | | | |
| Model 1 | Categorical cross entropy | *16* | *For the first 100 epochs:*<br>ADAM<br><br>*For the remaining epochs:*<br>Weighted Stochastic gradient descent (wSGD) with decoupled weight decay[58] for the remaining epochs (<br>Weight decay=1e-6,<br>Decay steps=1000,<br>Decay rate=0.5,<br>momentum=0.9). | 70 (200) | For ADAM:<br>1e-3<br>For wSGD:<br>3e-5 | 0.934 |
| Model 2 | | | | 49 (100) | | 0.887 |
| *GP4* | | | | | | |
| Model 1 (SCL) | Max. margin contrastive loss for SCL[53]<br><br>Binary cross entropy for classification | 16 | [Contrastive feature learning]<br>ADAM with<br>cosine decay for learning rate (decay step: 1000) 16<br><br>[Classification]<br>*For the first 100 epochs:*<br>ADAM 16<br><br>*For the remaining epochs:*<br>Weighted Stochastic gradient descent (wSGD) with decoupled weight decay[58] (Weight decay= 1e-4,<br>Decay steps=50, momentum=0.9).<br>We also applied exponential decay to the learning rate (Decay steps: 50, decay rate: 0.5). | 41 (200) $^{+}$ | [Contrastive feature learning]<br>1e-5<br><br>[Classification]<br>For ADAM:<br>1e-3<br>For wSGD:<br>3e-5 | 0.781$^{+}$ |
| Model 2 (SCL) | | | | 110 (200) $^{+}$ | | 0.816$^{+}$ |
| Model 3 | Categorical cross entropy | | | 90 (200) | | 0.801$^{+}$ |
| Model 4 | | | | 191 (400) $^{+}$ | | 0.805 $^{+}$ |
| Model 5 | | | | 393 (400) $^{+}$ | | 0.821$^{+}$ |
| Model 6 | | | | 298 (300) | | 0.758 |
| *GP5* | | | | | | |
| Model 1 | Categorical cross entropy | 32 | ADAM<br>A class weight was applied instead of oversampling approach. The class weight is the inverse class proportion.<br>0.7927398 for positive class (i.e., Gleason pattern 5)<br>0.2072602 for negative class | 192 (300) $^{+}$ | 1e-3 | 0.996 |
| Model 2 | | | | 289 (300) $^{+}$ | | 0.994 |
| *HGPIN* | Categorical cross entropy | *16* | ADAM | 31 (50) | 1e-3 | 0.982 |
| *Cribriform pattern* | Categorical cross entropy | 32 | ADAM | 476 (600) | 1e-3 | 0.943 |
| *Ductal morphology* | Categorical cross entropy | 64 | Stochastic Gradient decent with the cycle polynomial decay learning rate (Decay step: 12550, momentum=0.0, power=2, max. learning rate of 9.9725e-9) | 28 (200) | 1e-3 | 0.948 |
| *Blood vessel* | Categorical cross entropy | 128 | ADAM | 100 (200) | 1e-3 | 0.940 |
| *Nerves* | Categorical cross entropy | 128 | ADAM | 105 (200) | 1e-3 | 0.921 |
| *Inflammatory cell infiltration* | Categorical cross entropy | 128 | ADAM | 100 (100) | 1e-3 | 0.926 |

## Ensemble learning for Gleason pattern detection

Given that Gleason patterns are heterogenous and may have different latent appearance distribution, we applied ensemble learning to improve pattern detection accuracy and to increase the likelihood of having models robust to variation in appearance. Furthermore, we hypothesized that having multiple small models trained with different architectures and deep learning approaches provides better generalizability compared to a single small model.

### Gleason pattern 3

For Gleason 3, we trained two models with different model architecture configurations (**See Table 11**) on a dataset with magnification levels of 5×, 10×, and 20× and one model on a dataset with magnification levels of 10× and 20×. The rationale behind training on such datasets is to capture magnification-invariant features related to Gleason pattern 3; specifically, we were interested in well-differentiated glandular appearance and the connective tissue space between glandular structure as these features have an important role in discriminating Gleason pattern 3 from 4 or 5.

### Gleason pattern 4

Because Gleason pattern 4 has a broader tumor appearance compared to other Gleason patterns (i.e., 3 and 5), we used six PlexusNet models with different model configurations and training schemes: one model contained 3 transformer blocks after the last convolutional layer and before the full connection step, two models of two different epochs were trained using supervised contrastive learning[53]; two models of different epochs were trained in the same training setting and one model trained based on multi-task learning. The intuition behind considering models from different epochs or with different algorithms is to hopefully lessen the effect of the latent distribution shift that may occur during model training and to increase the likelihood of capturing informative features. In our case, we selected two models where the epoch distance between these models corresponded to the half distance to the last epoch with the best model (e.g., the best model found at 393 epochs and accordingly, we selected the next model at 191 epochs; both models were registered due to the reduction in loss values on the in-validation set compared to the prior epoch). the optimal hyperparameters were configured using the trial-and-error method for the detection of Gleason patterns.

### Gleason pattern 5

For Gleason pattern 5, we utilized two models with soft attention blocks that generate pixelwise attention maps and trained on datasets of TMA spot images with 10× magnification level. Our decision to use the pixelwise soft attention block was due to its superior performance for Gleason pattern 5 detection and histological characteristics of Gleason pattern 5 as part of neural architecture search. As demonstrated by **Figure 10**, histological characteristics of Gleason pattern 5 include tumor cells infiltrating into the connective tissues (captured by local attention) and glandular structure disappearance (captured by global attention).

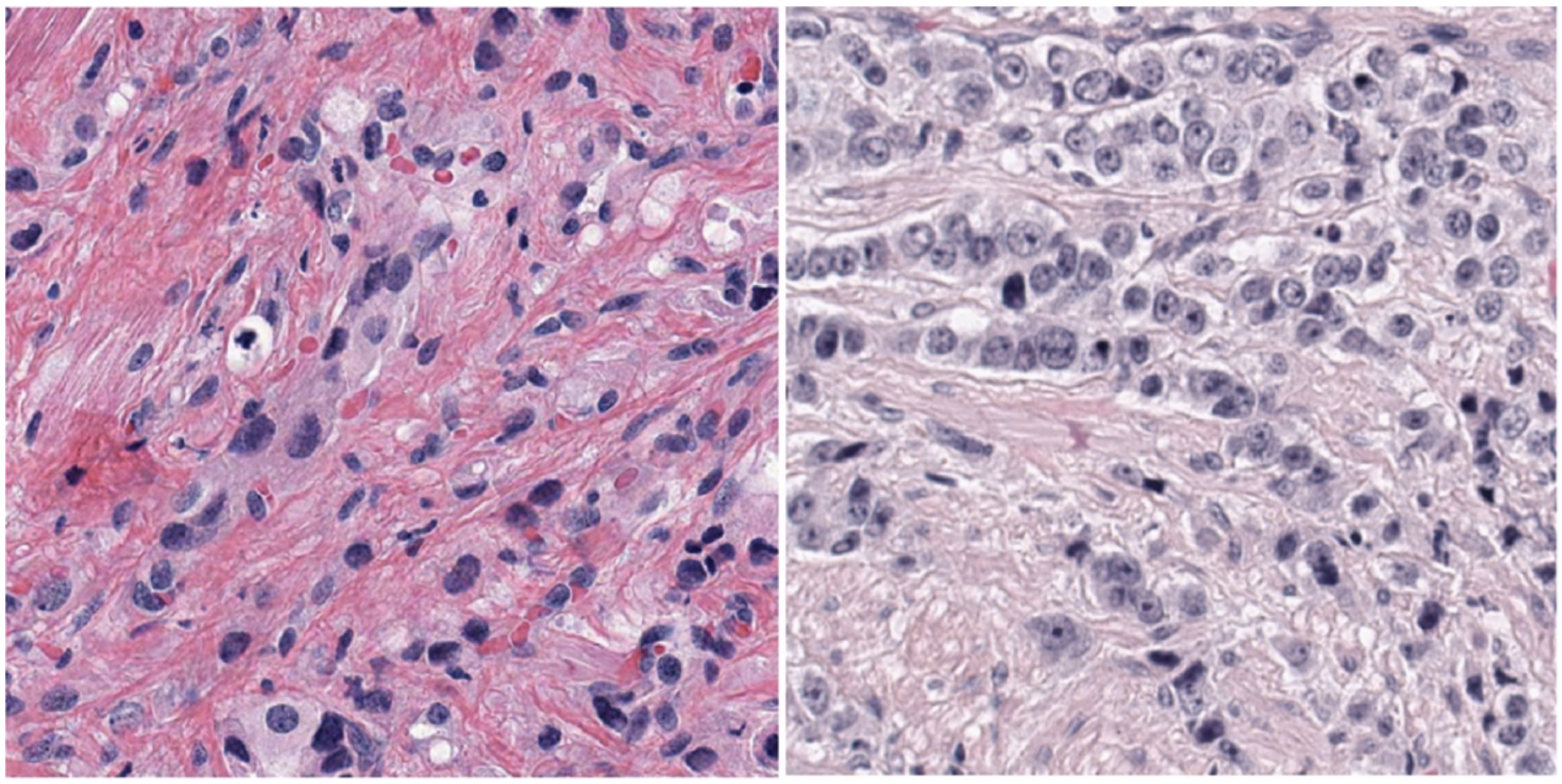

*Figure 10 illustrates two patches at 10× objective magnification for Gleason pattern 5 (Patch dimension: 512×512 px.).*

## Threshold selection

We applied the brute force approach to determine the threshold (operating point) with the best Cohen Kappa on the in-training validation set for patch-level classification of findings. Here, we increased the initial threshold 0.1 by 0.01 and measured the Cohen kappa for interrater agreement between the annotator and vPatho at the patch level (tiles); our approach can be considered similar to the Youden's J statistics where the largest J value is determined[59]. Bootstrapping with 10,000 times resampling was applied to determine the thresholds. **Figure 11** provides a code snippet for the threshold determination.

```python
import numpy as np
from sklearn.metrics import cohen_kappa_score

def boostrap(func, y_true, y_pred, n_bootstrap=10000,rng_seed=1234):
    bootstrapped_scores=[]
    rng = np.random.RandomState(rng_seed)
    for i in range(n_bootstrap):
        indices = rng.randint(0, len(y_pred),len(y_pred))
        score =func(np.array(y_true)[indices], np.array(y_pred)[indices])
        bootstrapped_scores.append(score)
    middle =np.median(bootstrapped_scores)
    return middle
thresholds = []
result_middle = []

for threshold_ in range(10,100,1):
    threshold = float(threshold_)/100
    m = boostrap(cohen_kappa_score, y_true, y_pred>threshold)
    thresholds.append(threshold)
    result_middle.append(m)
index_w_best_c=np.argmax(result_middle)
print(thresholds[index_w_best_c], result_middle[index_w_best_c])
```

*Figure 11 illustrates the code snippet. This code determines the threshold with the highest Cohen kappa score. "y_true": the ground truth labels of the patches from the optimization set and "y_pred" is the predicted confidence scores for these patches. We considered the median of the bootstrapped scores to mitigate the effects of outliers. After calculating the scores for all thresholds, the threshold with the highest Cohen Kappa "index_w_best_c" is determined. The thresholds are given in the supplementary Material and Method section.*

## Test Conditions

To cover the full range of PCa evaluation during routine clinical practice, we defined eleven test conditions that are important for pathology evaluation and reports. The subset for each test condition was defined prior to running the test (**Table 2** provides a summary of the datasets utilized for different test conditions). The eleven test conditions are as follows.

### Cancer Detection on H&E-stained Slides

We made a collection of histology images that included slides archived for approximately 20 years to determine the utilization boundary of our approach on slides affected by the aging process. Spot images of a tissue microarray (TMA) and whole mount images of prostatectomy specimens were also considered as they represent the breadth of tissue sampling for prostatic tissues.

Given that a single whole-mount (WM) image corresponds to roughly 30 stretched biopsy core images and the time- and labor-intensive effort of a high precision annotation of WM images for prostate cancer, we randomly selected 46 radical prostatectomy specimens with a total of 368 WM images [~11,040 biopsy core images with prostatic tissues or 894,240 patch images (512×512 pixels, 512 $\mu$m, at 10x by a pixel mapping of ~1 $\mu$m per pixel)] from 136 cases between 2016 and 2019.

A single pathologist (CK) annotated PCa areas on 368 slides. In contrast, each TMA spot was previously labelled for PCa presence based on the judgment of a single pathologist (RB) and whether it originated from cancer lesions in prostatectomy specimens. The cancer lesions in the historic McNeal slides were already delineated as part of a previous study investigating the tumor distribution in prostatectomy[32]. Having different sample collections is reflective of real-world experience in the clinic and research hospitals.

For the evaluation of agreement levels between automated and manual annotations of PCa, a 49% or above confidence score was used as the decision threshold for (integer) defining the presence of PCa in a patch. This threshold was deemed sufficient because of the well-calibrated DL model (**See Supplementary file 1**).

## Cancer Detection on each H&E-stained Slide

We measured the area under the receiver operating characteristic curve (AUROC) on every patch from every slide, using their predicted confidence scores and their true labels for PCa presence. To account for variation in number of patches between slides, we evaluated slide-level performance using the mean per-slide AUROC.

We also assessed Cohen's kappa[60] as a measurement of the degree of agreement between the model labels and the true labels made by the pathologist (CK) on image patches for each slide. Then, we determined the median per-slide Cohen kappa and its range. This test condition reproduced the spatial PCa detection (annotation) on each slide as well.

### Cancer Detection on H&E-stained Slides from a Single Case

This test task utilized all slides from the prostatectomy specimens of 46 cases with complete tumor annotation. We determined the average performance for cancer detection on patch images per case (per-case AUROC). This test condition reflects a clinical condition in pathology where we collectively evaluate all slides from a single case.

### Spot Detection of PCa on an H&E-stained TMA

This test task utilized H&E-stained histology images from 4 TMAs with a total of 1,129 spots from 339 prostatectomy specimens, which had on average three cores from each case and contained TMA cores with normal tissues. Each spot image represents a single tissue core (Diameter range: 0.6-1 mm). We determined the detection accuracy (i.e., sensitivity, specificity, negative and positive likelihoods, positive and negative predictive values and AUROC) for cancer detection on all spot images and the average AUROC after stratifying by TMA (per-TMA AUROC). This test condition represents a research condition where the TMA evaluation of cancer presence is needed or when we have a very limited tissue amount to investigate for cancer presence during the clinical routine or research. Parallelly, we measured the impact of overlapping of neighboring patches on the detection performance.

### PCa Detection on Old Slides with Weak H&E Staining due to the Aging Process

We randomly selected 13 H&E whole slide images originating from the historic McNeal dataset[32]. These H&E slides were archived over a period more than 20 years. From these slide images, we generated 11,862 patch images (512×512 pixels, 512 $\mu$m, at 10× by a pixel mapping of 1 $\mu$m per pixel), including 3,552 patch images with PCa. We applied a reference-free version of Macenko's stain normalization algorithm for color intensity optimization of the patch images [43] (See Figure 1A.1).

For accuracy evaluation, we measured the area under the receiver operating characteristic curve (AUROC) on all patch images using their predicted confidence scores and their true labels for PCa presence. This test condition replicates the challenge of cancer detection on images originating from slides with faded staining to assess the limitations of our approach.

### Tumor Volume Estimation

We measured the association between predicted and true tumor volume on 46 prostatectomy specimens with complete delineation of cancer lesions (368 slides). We utilized the well-established grid method, that described in detail in our previous study[4]. In brief, we define a two-dimensional space for each slide where a single patch image corresponds to a pixel (a pixel corresponded to 512×512 $\mu$m dimension of a slide image at 10x objective magnification level). Then, we count the total pixels (patch images) that are positive for prostate cancer. We also determine the background tissue using image thresholding (Otsu's method) and exclude the white areas located outside the background tissues in all slides for each case. After that, we determined the number of patch images generatable from the background, which also affects the number of pixels. Finally, the total number of pixels with PC is divided by the total number of pixels of the background tissues to estimate the tumor volume (TuVol%) in each case (**Figure 1B**). The coefficient of the regression score determined the correlation of TuVol% between the ground truth and the AI solution at the case level. The pairwise Welch's t-test was applied to identify the significance of the variation between the ground truth and the AI solution for tumor volume estimation[61].

### Gleason Patterns and ISUP Grading

This test condition defined three tasks. The first task is to determine Gleason patterns on very limited tissues (~256 – 512 $\mu$m) suitable for laser capture microdissection[62]; the second and third tasks are to determine ISUP grade on biopsy cores and prostatectomy specimens respectively.

The ISUP grading system for each biopsy core is defined by the most frequent Gleason pattern for the primary Gleason pattern and the highest Gleason pattern for the secondary Gleason pattern. In contrast, the ISUP grading system for radical prostatectomy specimens considers the first and second most frequent Gleason patterns cross the specimen. The second most frequent Gleason pattern is qualified for the secondary Gleason pattern if it is equal to or exceeds 5%. If the second most frequent Gleason pattern is below 5%, the secondary Gleason pattern will be the same as the primary Gleason pattern and the second most frequent Gleason pattern will be the tertiary Gleason pattern.

Since there are no widely agreed rules for limited tissues to report ISUP grading and the patch level is the smallest unit required to assess ISUP grading at slide and case level, we focused on evaluating the accuracy for the Gleason pattern detection in such limited tissues. The decision thresholds of 35%, 65% and 93% for GP3, GP4 and GP5, retrospectively, were determined on the development set using a brute force algorithm that identified thresholds with the best interrater reliability (Cohen's Kappa[60]). After fixing these thresholds, Gleason pattern presence was determined in patch images to measure inter-rater agreement.

Two test conditions assessed the concordance level between pathologists and vPatho for Gleason grading on the ISUP dataset (biopsy cores) and the Stanford's external dataset (radical prostatectomy. We aimed to use the same vPatho for these two different grading conditions to identify histopathological factors impacting the concordance level of the current ISUP grading system on radical prostatectomy specimens. We considered both 10x and 20x magnification levels to determine the magnification level with a better Gleason pattern detection.

### Very limited tissue samples for laser microdissection

Non-overlapping patch images covering GP3, GP4 and GP5 were available to simulate a Gleason pattern detection on tissue dimension suitable for laser capture microdissection (**Table 12**). These images were generated from 47 prostate cancer regions randomly selected in 60 whole-mount slides of 10 radical prostatectomies (average 4.7 regions per radical prostatectomy). Each patch image included only one Gleason pattern thanks to a time-intensive effort to annotate highly homogenous regions for each pattern. The label of patch images is defined based on the annotation data curated by a single pathologist "YT". A patch image is positive for one of the Gleason patterns, when at least 10% of the patch image is positive. For each Gleason pattern, we measured the detection accuracy and inter-rater agreement between the pathologist and the DL models (See evaluation metrics). Finally, we repeated the evaluation at 20x to evaluate the agreement level between DL models and the single pathologist (**Table 12**).

*Table 12 describes the numbers of patch images for each finding considered to evaluate our deep learning models for Gleason pattern detection on very limited tissue samples. Manual quality control to ensure that these patches include only a single finding was performed.*

| Finding | *Number of patch images (%)* | |
|---|---|---|
| | 10×objective magnification | 20× objective magnification |
| High-grade Prostatic intraepithelial neoplasia | 250 (22.2) | 845 (22.0) |
| Gleason pattern 3 | 311 (27.6) | 1211 (31.5) |
| Gleason pattern 4 | 454 (40.2) | 1374 (35.8) |
| *Cribriform glands* | 36 (7.9) | 117 (8.5) |
| *Poorly formed glands* | 210 (46.3) | 538 (39.2) |
| *Fused glands* | 19 (4.2) | 57 (4.1) |
| *Poorly formed and fused glands* | 189 (41.6) | 662 (48.2) |
| Gleason pattern 5 (Mixed) | 113 (10.0) | 410 (10.7) |

## Biopsy cores

We considered the reference image database for Gleason patterns (GPs) and Gleason grading provided by ISUP; an international organization responsible for the histopathological definition of Gleason grading [35]. These reference images were graded by the ISUP member team according to majority rule[35]. This database represents a unique and independent resource to evaluate agreement level between the expert panel and DL models. To determine ISUP grades on biopsy cores, we considered reference images captured at 20× objective magnification. By considering different sizes of these images, we generated patches (512×512 $\mu$m; 1 pixel = 1 $\mu$m) for each ISUP image at 10× objective magnification after downsizing the original histology images or (~256×256 $\mu$m; 1 pixel = 0.5 $\mu$m) at 20x objective magnification. We populated the patch number for each ISUP images by applying 50% overlap between patches to increase the likelihood of a complete appearance of Gleason patterns in these patches with effective computational workloads. Then, we counted the positive patches for each GP. After implementing a conditional algorithm according to the ISUP grading system for the biopsy core, we estimated the ISUP grade for each biopsy core and evaluated the agreement level for tumor grading between the expert panel and our approach.

### Radical prostatectomy

A total of 136 patients who underwent radical prostatectomy were available for this task. These prostatectomy specimens were already evaluated during routine clinical practice by six different board-certified pathologists between 2016 and 2019. This cohort reflects cases seen during routine clinical practice. Using the chart review, we acquired pathological and clinical information for each case. The information included the tumor stage, the status of locoregional lymph node metastases, surgical margin status, the year of the pathology report, the ISUP grade and the pathologist who conducted the pathology evaluation.

After excluding the white background, all whole mount slide images were tiled into 512×512 pixels patches (~512 $\mu$m). Then, we counted the number of positive patches for each GP. Finally, we developed and applied a conditional algorithm to grade PCa in each case according to the ISUP grading system for prostatectomy. The patch images were not overlapped because of the high dimension of WM images compared to biopsy cores.

We measured the agreement level between a pool of 6 pathologists who evaluated these cases during the clinical routine and our approach. We repeated our agreement evaluation after we adjusted the decision threshold from 5% to 10% for secondary and tertiary Gleason patterns since our previous study showed that eyeball judgement can underestimate the tumor size by 50% of the original size[4].

### Sorting the Slides According to the Cancer Presence Status

We wanted to determine the sorting accuracy of slides with PCa since this measurement is relevant to improving clinical workflows. Therefore, we considered different types of slides and datasets to verify the model performance in identifying slides with PCa (**Table 2**). These slides originated from different institutions with different protocols for tissue preparation or different backgrounds (i.e., lymph node tissues). Here, two positive patches (in which the tumor probability exceeds the threshold of 49%) were enough to mark the slide as positive**.** We assessed the confusion matrix and measured the true positive rate (TPR) or positive predictive value (PPV), as well as the true negative rate (TNR) or negative predictive value (NPV), to objectively evaluate the sorting accuracy.

### Detection of ductal morphology

A total of 1,247 non-overlapping patch images (512×512 pixels ~ 512×512 $\mu$m at 10× objective magnification) covering ductal adenocarcinoma were available to simulate a ductal morphology detection. These patches were generated from 38 prostate cancer regions randomly selected from 2 whole-mount slides of 2 radical prostatectomies (**Table 13)**. Since ductal adenocarcinoma is a rare cancer morphology (approximately 0.4% to 0.8 % of radical prostatectomies), we weighted the patch number in favor of ductal adenocarcinoma for a balanced performance evaluation. The label of patch images is defined based on the annotation data curated by a single pathologist "YT". Each patch image was covering a single entity thanks to a time-intensive effort to annotate highly homogenous regions. Given that cancer lesions with ductal morphology may incorporate empty spaces, a patch image was considered positive for the ductal morphology, when at least 40% of the patch image was positive and the white background area in the patch image was not more than 60%. For the ductal morphology, we measured the detection accuracy and the inter-rater agreement between the pathologist and the DL model.

The definition of ductal morphology primarily depends on the ductal appearance of prostate cancer and splitting a region with ductal adenocarcinoma of prostate into patches causes the disappearance of ductal morphology in some patches. Therefore, we specifically reviewed and included only patches that represent their corresponding finding labels. The threshold was identified and set to 90% using the development set, using the same approach for the threshold determination for the Gleason patterns detection already described in the earlier section.

*Table 13 describes the numbers of patch images to evaluate the detection model for ductal adenocarcinoma.*

| Finding | *Number of patch images (%)* |
|---|---|
| Ductal adenocarcinoma | 1247 (59.0) |
| Non-ductal prostate cancers | 865 (41.0) |
| -Gleason pattern 3 | 315 (36.4) |
| -Gleason pattern 4 | 435 (50.3) |
| -Gleason pattern 5 | 115 (13.3) |

## Detection of cribriform pattern

A total of 92 non-overlapping patch images (512×512 pixels; ~512×512 $\mu$m at 10× objective magnification) covering cribriform Gleason pattern 4 (n=37) or ductal adenocarcinoma with cribriform architecture (n=58) were available to simulate a cribriform pattern detection (**Table 14**). The true label of patch images is defined based on the regional delineation and its labels on histology slides made by a single pathologist “YT”. Each region had a single finding to achieve more homogenous dataset to simulate the detection performance for cribriform patterns. A patch image was considered positive for the cribriform pattern when at least 50% of the patch image was positive. As a negative control, we considered a total of 836 covering Gleason patterns 3 to 5 with exception of Gleason pattern 4 with cribriform patterns. These images for negative control were originated from the previous test condition for Gleason pattern detection.

For the cribriform pattern, we measured the detection accuracy and the inter-rater agreement between the pathologist and the DL model. Given that the definition of cribriform pattern primarily depends on the distinctive appearance of holes within the cancer cell aggregation, we specifically reviewed these patches to ensure that the positive patches included the distinctive appearance of the hole in prostate cancer prior running the evaluation. The threshold was identified and set to 65% on the development set using the same approach for the threshold determination for the Gleason patterns detection described earlier.

*Table 14 describes the numbers of patch images to evaluate the detection model for ductal adenocarcinoma.*

| Cribriform patterns | *Number of patch images (%)* |
|---|---|
| Yes | 92 (9.9) |
| No | 836 (90.1) |

## Detection of nerves

A single pathologist randomly delineated 158 nerves structures on the same 60 slides of 10 cases, resulting in a total of 739 patch images with nerve components (nerves and ganglions). Further, we considered 541 patches with Gleason patterns as negative group as these images did not include nerve structures. A patch image was considered positive for the nerve structures, when at least 5% of the patch image was positive to incorporate small nerve structures (~25.6 $\mu$m). Finally, we measured the detection accuracy and the inter-rater agreement between the pathologist and vPatho. Given the heterogeneity of prostatectomy specimens, we specifically reviewed these patches to ensure that these patches represent their corresponding finding labels prior the evaluation. A probability threshold of 35% was identified on the development set and fixed for testing using the same approach for the threshold determination described earlier.

### Detection of vessels

We generated a total of 1,455 non-overlapping patch images (512×512 pixels; 512×512 $\mu$m at 10× objective magnification) covering the general appearance of blood vessels to simulate a blood detection task. These images were generated from 150 blood vessels in various sizes randomly selected and annotated by a single pathologist on the previous 60 whole-mount slides of 10 radical prostatectomies. A patch image was considered positive for the blood vessel, when at least 5% of the patch image was positive (A threshold of 5% was considered to cover small vessels). The negative group included 3,153 patches from 50 cancer regions that were already prepared in the previous steps. Finally, the detection accuracy and the inter-rater agreement between the pathologist and the DL model were calculated. A probability threshold to classify the patches was set to 50%.

### Detection of inflammatory cell infiltration

We curated 268 non-overlapping patches with inflammatory cell infiltration (512×512 pixels; 512×512 $\mu$m at 10x objective magnification) to simulate the detection task of tissue regions with infiltrated inflammatory cells. Parallelly, a negative group with 500 patch images positive for cancers was randomly generated from 216 regions. A patch image was considered positive for inflammatory cell infiltration, when at least 3% of the patch image was positive since an inflammatory cell infiltration can be small (~10 $\mu$m).

Thereafter, we measured the detection accuracy and the inter-rater agreement between the pathologist and the DL model. Given the heterogeneity of prostatectomy specimens, we specifically reviewed these patches to ensure that these patches represent their corresponding finding labels. We used a probability threshold of 70.5% to classify the patches which is determined on the development set using the same approach described earlier.

### Detection of tumor precursors

A single pathologist screened for high-grade prostatic intraepithelial neoplasia lesions (HGPIN) in 60 whole-mount slides of 10 radical prostatectomy cases and identified 32 regions with HGIPN. From these regions, we curated 250 patches to examine the detection performance for HGPIN with 50% overlap (512×512 pixels; 512×512 $\mu$m at 10× objective magnification) to increase the likelihood of having HGPIN appearance in these patches. During generating the patches, we manually excluded the lumen patches after running an algorithm that identified and excluded lumen patches as background. A single patch is positive when 10% of the patch is positive for HGPIN (this threshold is arbitrary). As negative control, we used the patches with benign prostatic hyperplasia (BPH) and intraductal adenocarcinoma (IDC) as IDC is considered as the differential diagnosis of HGPIN whereas BPH and HGPIN are benign tumor-like lesions in the prostate [57,63-65]. During generating the patches, we manually excluded the lumen patches after running an algorithm that identified and flagged lumen patches as background. The test condition shall cover the detection of HGPIN in limited tissues where capturing HGPIN lesions require the utilization of laser microdissection. Identifying regions suitable for the laser microdissection is essential for studies investigating HGPIN lesions as well. We emphasize the readers that our sample size for HGPIN is adequate given that HGPIN are mostly limited and very small lesions found in up to 50% of prostatectomy specimens[66] and therefore preferably sampled using the laser microdissection[67]. A probability threshold to classify the patch according to HGPIN presence was fixed to 65%, which was identified using the brute-force search approach on the development set as described in the supplementary model development section.

### Reporting the Existence of Relevant Findings in a Case

We aimed to evaluate the feasibility to integrate the results from deep learning models into the pathology report. Here, we aimed to answer two major questions:

1) How integrable are the results provided by vPatho into the pathology report for prostatectomy specimens?
2) What information needed to be considered when implementing such results in pathology reports?

To answer these questions, we selected 136 patients who underwent radical prostatectomy and reflected the cases seen during routine clinical practice. Since these prostatectomy specimens were already evaluated during routine clinical practice and the evaluation results were documented, we reviewed the pathology reports for the ISUP grades, the tertiary Gleason patterns, cribriform pattern, and ductal morphologies for these cases before running the analyses.

## Investigating variables indicating the grade disagreement on prostatectomy specimens

The following analyses shall answer the question whether the current practice for ISUP grading on prostatectomy specimens has a direct effect on the grade disagreement (discordance) after controlling for other cofounders. We further wished to identify other independent factors associated with grade disagreement on prostatectomy specimens.

Our study involved experienced genitourinary pathologists from high-volume centers (>200 radical prostatectomy cases per year) to compare our AI solutions. A survey found a significant correlation between the number of biopsies evaluated by the pathologist per week and the grade concordance level with expert review[68]. Goodman et al concluded that the likelihood of discordance between pathology reports and expert-assigned Gleason scores is particularly elevated for small community hospitals compared to high-volume hospitals[69]. Therefore, defining the expertise based on the case volume instead of experience years (which is associated with a physician group who are at higher risk for medical errors due to their age [70]) is the most appropriate reflection of the clinical experience.

We investigated variables for their effects on the grade disagreement between the pathology reports and our approach using the binomial generalized linear mixed-effects model[71]. The variables were the pathologists who conducted the pathology evaluation, TuVol% made by AI and pathologists, the proportion of positive slides (which represents the vertical tumor extent), the cancer proximity to the prostatic capsule and grades (i.e., the ISUP grade made by pathologist and by AI-assisted approach), the year of tumor grading, and the total number of slides for each case. The cancer proximity to the prostatic capsule was determined using a mask function that considers only cancer areas that are within a border zone (zone 1) covering 10% of the prostate slice as **Figure 12** schematically illustrates and the registration of cancer presence was performed according to Eminaga et al [72].

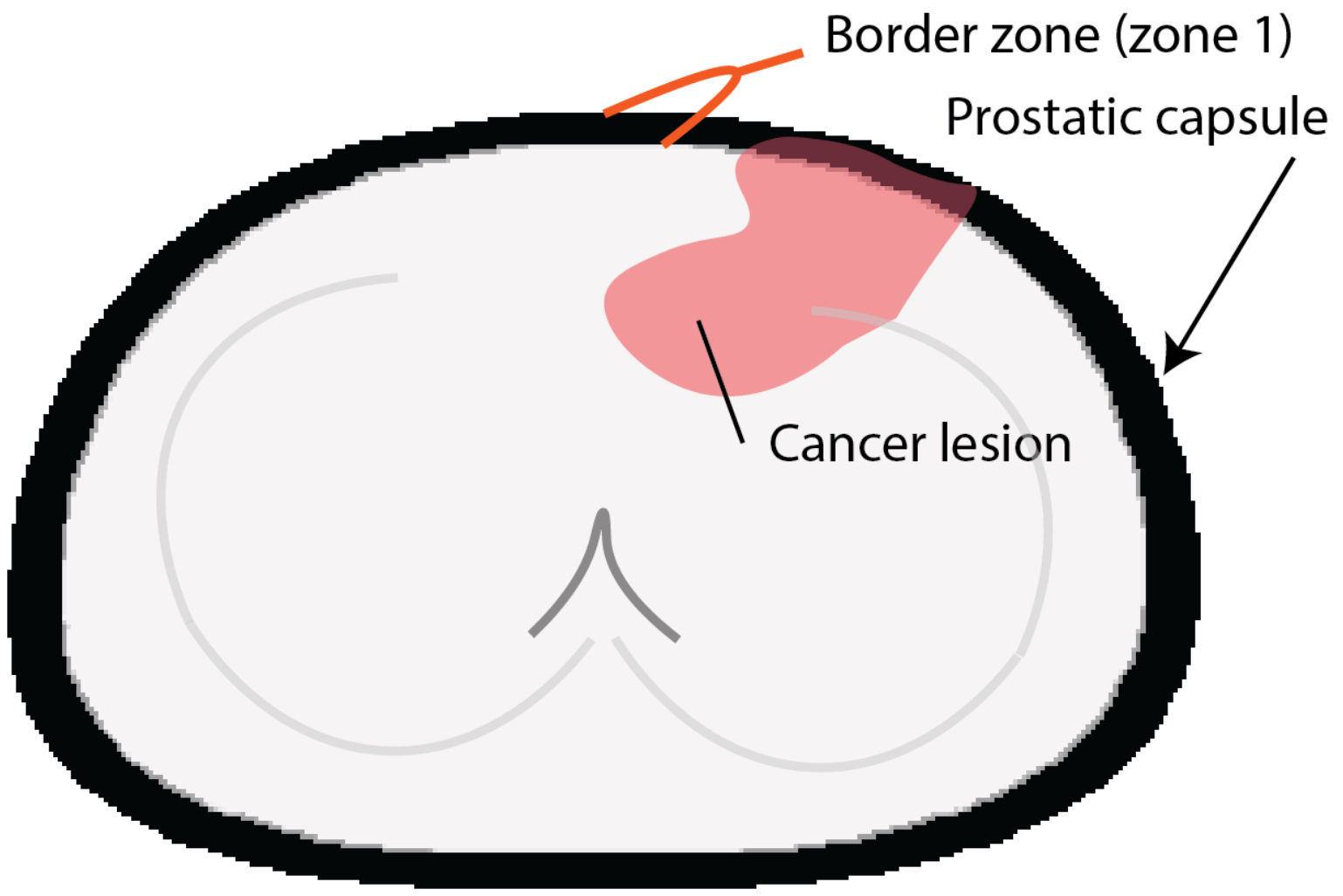

*Figure 12 schematically illustrates the definition of the border zone (zone 1) which makes 10% of the prostatic slice area and adjacent to the prostatic capsule.*

The year of tumor grading was incorporated as representation for variation in staining protocol over the time and adaptation period of the recent version of ISUP grading system in 2016. As model input, the standard scores were estimated for all variables listed above. As model output, we calculated the odd ratio and its 95% Wald Confidence Interval for each variable.

We designed four mixed-effects models to identify the effects of these variables on the grade disagreement with four goals:

1. To evaluate the indicative effect of the ISUP grading made by the pathologists on the grade disagreement regardless of the pathologist who provided the tumor grading.
2. To evaluate the indicative effect of the pathologist on the grade disagreement regardless of the ISUP grade made by the pathologists
3. To evaluate the indicative effects of variables that are determined by AI on the grade disagreement regardless of the pathologist who conducted the tumor grading that we compared with.

4. To evaluate the indicative effects of significant variables on disagreement while considering cancer proximity to the prostatic capsule, the tumor stage, and the status for locoregional lymph node metastasis status of prostate cancer.

The hypotheses for each model were:

H0: There is no relationship between the variables and the grade disagreement.

H1: There is a relationship between one or more variables and the grade disagreement.

Since multicollinearity may impact the stability of the generalized linear mixed-effects model, we assessed the variance inflation factor (VIF) of these variables to measure the multicollinearity in each model. Here, we considered a VIF below 2 as negligible collinearity [29]. The intraclass correlation coefficient was provided for the random effect to identify the proportion of variance that can be explained by the random effects[73].

We further evaluated the impact of a variable on the goodness of fit of each model by comparing Akaike information criterion (AIC)[74] of the model before and after excluding the variable; the analysis of deviance tables was computed to determine the comparison significance between a model with the variable and a model without the variable. The false discovery rate (used to correct P values) was estimated to determine the significance of the change in AIC.

We run a Monte Carlo Simulation (1,000 simulations) to measure the Power of the effects for all significant variables after setting the rate of the error type I (falsely rejecting the hypothesis -H0- that there is no relationship) to 10% (or 5% for each side). The function “powerSim” from “simr” package was utilized to measure the power for mixed-effects models[75].

Given that there is a collinearity in ISUP grades between AI and pathologists, we considered only the ISUP grades made by pathologists for goal 1, 2 and 4, and the ISUP grades made by AI for the goal 3. In goal 4, we furthermore repeated our model evaluation after replacing ISUP grades made by pathologist with those made by AI. **Table 15** reveals the variables for fixed effects and random effect. The "glmer" function from lmer4 package was used for the binomial logistic linear mixed-effects modelling with the default configuration [76].

*Table 15 describes the variables included for each mixed-effect model.*

| | Variables | |
|---|---|---|
| Model (Goal) | Fixed effects | Random effect (random intercept) |
| A | -Tumor volume in percentage<br>-Year of the pathology report<br>-Proportion of positive slides<br>-The total number of slides<br>-Gleason grade made by the pathologists | Pathologists |
| B | -Tumor volume in percentage<br>-Year of the pathology report<br>-Proportion of positive slides<br>-The total number of slides<br>-Pathologists | Gleason grade made by the pathologists |
| C | -Tumor volume in percentage<br>-Year of the pathology report<br>-Proportion of positive slides<br>-Gleason grade made by AI | Pathologists |
| D | -Tumor stage (pT)<br>-locoregional lymph node metastases status (pN)<br>-cancer proximity to prostate capsule<br>-Gleason grade made by the pathologists/by AI<br>-Tumor volume in percentage<br>-Proportion of positive slides | Pathologists |

The general equation for the mixed-effect model is

1. $y = X\beta + Zv + \epsilon$

Where y is a vector of the grade disagreement status, $X$ is a matrix of fixed variables and $\beta$ is a vector of fixed effects regression coefficient, $Z$ is a design matrix of the random effect, $\upsilon$ is a vector of the random effect and $\epsilon$ is a vector of the residual.

Finally, we conducted model-based causal mediation analyses to measure the indirect causal mediation effect of the pathologist or another significant variable (other than ISUP grading) on the grade disagreement. To simplify the mediation analyses, we defined low-grade and high-grade groups based on ISUP grades (ISUP grades 1 and 2 vs. ISUP grades 3-5) as a binary group instead of considering the five ISUP grades.

We further evaluated the variables for their association with the mediators. When we investigated the pathologist factor as mediator, we selected variables that are defined based on the pathologist's decision (tumor stage, pN status, and ISUP grading). Furthermore, we incorporated TuVol%, the proportion of positive slides and the year of tumor grading as these factors may affect the finding search on WM histology slides. Such associations were assumed when the P value is $<0.2$ (two-tailed test).

A sensitivity analysis with 1,000 bootstrap resampling was carried out to validate the results for the causal mediation effects after iterating the mediator coefficient with different constants[77].

The mediation and sensitivity analyses were conducted using the package "mediation"[77]; the moderation of the relationship between the independent variables and the mediator was realized using a linear model, whereas the overall effect was estimated using a binomial logistic regression model[78]. The results for mediation analyses were presented in flowchart diagrams.

The contingency table was evaluated using Pearson's chi-squared test[79]. The medians were evaluated for their difference between the groups using Mann Whitney

Wilcoxon test[80] and the means between the groups using Welch's t test[61]. A significant difference was determined when the two-tailed test showed a P value ≤ 0.05. The 95% Clopper-Pearson confidence interval for the proportion was estimated[81].

## Evaluation Metrics

The discrimination accuracy for each endpoint was evaluated using AUROC. The AUROC reveals the classification performance at different thresholds; a higher AUROC indicates a better classification accuracy, where an AUROC of 1 represents the highest accuracy [82]. Using the thresholds determined on the development set, we calculated the sensitivity, specificity, negative and positive likelihoods. Furthermore, we estimated the positive and negative predictive values for different prevalence rates (e.g., patch proportion with prostate cancers). The agreement rate was measured using Cohen kappa for binary category or weighted quadratic kappa for more than 3 categories. The agreement rate was described according to Cohen J., who introduced Cohen kappa and weighted kappa (**Table 16**) [60,83].

*Table 16 lists the kappa values and the corresponding agreement levels.*

| Kappa | Agreement |
|---|---|
| <0.0 | Less than chance agreement |
| 0.01 – 0.20 | Slight agreement |
| 0.21 – 0.40 | Fair agreement |
| 0.41 – 0.60 | Moderate agreement |
| 0.61 – 0.80 | Substantial agreement |
| 0.81 – 0.99 | Almost perfect agreement |

The coefficient of the regression score determined the correlation of relative tumor volumes between the ground truth and the AI solution at the case level. The pairwise Welch's t-test was applied to identify the significance of the variation between the ground truth and the AI solution for tumor volume [61]. The reported P values are two- tailed, and statistical significance was considered when $P \leq 0.05$. The uncertainty measurement (95% CI) for AUROC, Cohen kappa and weighted kappa was determined by bootstrapping with 100,000 replications[84]. The Cooper-Pearson interval was utilized to calculate the 95% confidence interval[81] for detection accuracies (i.e., sensitivity, specificity, positive and negative likelihoods, positive and negative predictive values).

## Software and Hardware settings

Our analyses were performed using Python 3.6 (Python Software Foundation, Wilmington, DE) and R 3.5.1 (R Foundation for Statistical Computing, Vienna, Austria). We applied the Keras library, a high-level wrapper of the TensorFlow framework, to develop the models. All analyses were performed on a GPU machine with a 32-core AMD processor with 128 GB RAM (Advanced Micro Devices, Santa Clara, CA), 2 TB PCIe flash memory, 5 TB SDD hard disks, and a single NVIDIA Titan V GPU with 12 GB VRAM.

## Author contributions

OE, MA, and OB designed the study. CK, RN, JB, RF, RB, and OE collected the data. MA, OE (development set curation team) RN, YT, RB and CK (test set curation team, a single pathologist was involved in each task) made the annotations. OE developed cMDX, performed data modeling and analyzed the data. YH assisted OE during the model development. OE and YH developed viewer tools. JL reviewed the results from statistical perspective. OE, JB, RB, OB, and MA drafted the manuscript. All authors read and approved the manuscript.

## Funding
None.

## Notes
**Role of the funder**
No involvement in the study design nor execution.

## Disclosures

Nothing to disclose.

## Data Availability

Stanford's datasets are not publicly available due to patient privacy regulations and the internal data sharing policy. However, other public datasets are available, and their sources are given in "image database" section (Supplementary material). A cMDX file and the cMDX viewer tool are available for demonstration purposes (https://github.com/oeminaga/cmdx_report.git), and a python package for the PlexusNet architecture is publicly available (https://github.com/oeminaga/PlexusNet.git).

## Code statement

An abstract version of the pipeline will be provided upon acceptance.